\documentclass[aps,preprint,nofootinbib,eqsecnum]{revtex4}
\usepackage{amsmath}
\usepackage{amsfonts}
\usepackage{amssymb}
\usepackage{graphicx}
\usepackage{amscd}

\newcommand{\cL}{{\mathcal{L}_0}}

 \allowdisplaybreaks

\begin{document}

\thispagestyle{empty}
\begin{flushleft}
USCHEP/0302ib1
\hfill hep-th/0302151 \\
UT-03-04\hfill February 2003 \\
\end{flushleft}

\vskip 1.5 cm
\bigskip

\begin{center}
 \noindent{\Large \textbf{Analytic Study of Nonperturbative }}\\
 \noindent{\Large \textbf{Solutions in Open String Field Theory}}
 \noindent{
 }\\
\renewcommand{\thefootnote}{\fnsymbol{footnote}}

\vskip 2cm
{\large
I. Bars$^{a,}$\footnote{e-mail address: bars@usc.edu},
I. Kishimoto$^{b,}$\footnote{e-mail address:
 ikishimo@hep-th.phys.s.u-tokyo.ac.jp} and
Y. Matsuo$^{b,}$\footnote{e-mail address:
 matsuo@phys.s.u-tokyo.ac.jp} } \\
{\it
\noindent{ \bigskip }\\
$^{a)}$ Department of Physics and Astronomy,\\
University of Southern California, Los Angeles, CA 90089-0484, USA \\
\noindent{\smallskip  }\\
$^{b)}$ Department of Physics, Faculty of Science, University of Tokyo \\
Hongo 7-3-1, Bunkyo-ku, Tokyo 113-0033, Japan\\
\noindent{ \smallskip }\\
}
\bigskip
\end{center}
\begin{abstract}
We propose an analytic framework to study the nonperturbative
solutions of Witten's open string field theory.
The method is based on the Moyal star formulation
where the kinetic term can be split into two parts.
The first one describes the spectrum of two identical half strings
which are independent from each other.
The second one, which we call midpoint correction,
shifts the half string spectrum to that of
the standard open string.  We show that the nonlinear equation
of motion of string field theory is exactly solvable
at zeroth order in the midpoint correction.
An infinite number of solutions are classified in
terms of projection operators. Among them,
there exists only one stable solution which is identical to
the standard butterfly state.
We include the effect of the midpoint
correction around each exact zeroth order
solution as a perturbation expansion which can be
formally summed to the complete exact solution.
\end{abstract}
\vfill
\maketitle
 \vfill\setcounter{footnote}{0} \renewcommand{\thefootnote}{%
 \arabic{footnote}} \newpage

\section{Introduction}


{}Starting with the original work of Witten \cite{Witten} string field
theory has been strongly tied with noncommutative geometry. This is in the
spirit of treating the open string field as an infinite dimensional matrix,
while the definition of the star product is formally identical to matrix
multiplication. This viewpoint has been pursued in more detail by explicitly
splitting the left and right degrees of freedom in the split string
formalism \cite{splitSFT}.

Recently, this formal correspondence played a major r\^{o}le in importing
basic ideas of noncommutative geometry to string field theory. One of the
stimulating ideas is the vacuum string field theory (VSFT) proposal \cite%
{VSFT}. With an assumption on a simplified kinetic term to describe the
tachyon vacuum, the classical solutions that would describe the D-brane are
given by the noncommutative soliton \cite{GMS} (=projector). It is well
known that projectors are the fundamental geometrical objects in
noncommutative geometry since they represent the K-homology group. The proof
of the VSFT conjecture on the kinetic term was, however, difficult and there
remained many open questions.

The Moyal formulation \cite{B}\cite{BM1}\cite{BM2}\cite{BKM1}\cite{B2}
(MSFT) is an explicit representation of Witten's string field theory in
terms of the Moyal product, which is the main language in noncommutative
geometry. Unlike previous proposals of the split string field framework,
particular attention was paid to solve the ambiguity at the midpoint. In the
context of MSFT, the subtlety is reflected in the form of the associativity
anomaly \cite{BM1} of the infinite dimensional matrices $T,R,v,w$ which
provide the change of variables from the open string coordinates to the
canonical pairs of the Moyal product. The anomaly is resolved by deforming
the non-associative algebra among $T,R,v,w$ to an associative one, by
introducing a regulated version of these matrices. All the elements of
string field theory such as, perturbative spectrum, Neumann matrices and
Feynman rules, were explicitly written in terms of the regularized
framework, and their equivalence to other frameworks, when the regulator is
removed, was demonstrated in \cite{BM2}\cite{BKM1}. Other proposals of the
Moyal formalism \cite{DLMZ}\cite{Moyal} are equivalent to the original one
\cite{B2}.

In this paper, we take a step toward the classification of
nonperturbative solutions of string field theory. The method is
based on the splitting of the kinetic term into two parts. The
first one gives a description of the open string where the left
and right half strings completely decouple. The second term gives
the correction to the split string description to recover the
correct spectrum of the open string. We call it the ``midpoint
correction'' since it carries the information that two half
strings are indeed connected at the midpoint.

We will show that there exists a basis which diagonalizes the first part of
the kinetic term and the nonlinear interaction term at the same time. The
combined nonlinear system (namely Witten's action without the mid-point
correction of the kinetic term) becomes a completely solvable matrix model.
We can obtain all the exact solutions that are invariant under translations
by using the projection operators under the star product. These provide a
basis for classifying all nonperturbative solutions of open string field
theory on the D25 brane. We derive the spectrum at zeroth order in the
midpoint correction and then compute the effect of the midpoint as a
perturbation.

The midpoint correction has been neglected in the literature because the
coefficients in front of it seems to disappear in the naive limit of the
regularization. It has, however, a finite contribution to the spectrum and
cannot be discarded. In this paper, we propose to treat it as a correction
to any of the zeroth order exact solutions, and provide a formal expression
that sums up the perturbation series to the full exact solutions.

The situation turns out to be similar to the discussions of
noncommutative scalar field theory \cite{GMS}. There, in large
$\theta $ (noncommutativity parameter) limit, the potential term
gives the dominant contribution, and the equation of motion is
solved by projectors. For finite $\theta $, the
kinetic term is incorporated as a perturbation \cite{GMS}\cite%
{soliton-scattering}\cite{soliton-stability} which gives corrections to the
noncommutative soliton. In string field theory, usually there is no such
parameter $\theta $ which can be adjusted to justify the validity of the
perturbation expansion. However, in the Moyal formulation with a regulator,
we have some freedom to choose the string oscillator frequencies (denoted as
$\kappa _{n}$), while keeping the basic algebraic structure of string field
theory, including important relations such as the nonlinear relations of
Neumann coefficients derived by Gross and Jevicki \cite{GJ}. While these
parameters get fixed to the usual ones $\kappa _{n}=n$ at the end, we may
use the behavior of the theory as a function of this degree of freedom to
define the perturbation theory in the intermediate steps. In section \ref%
{perturbation}, we give a construction of the exact solution of the full
theory as the expansion of the midpoint correction along this idea. The
perturbative expansion can be determined uniquely at each order with a
condition (relating to the stability) on the spectrum of the open string
solution. The perturbation series can then be summed up to a formal
expression that represents the full exact solution.

\section{The setup}

\subsection{Moyal star formulation}

The starting point of our discussion is Witten's action in the Siegel gauge
written in the Moyal star formulation of String Field Theory (MSFT),
\begin{equation}
S\left( A\right) =-\int d^{d}\bar{x}\,Tr\left( \frac{1}{2\alpha ^{\prime }}%
A\star (L_{0}-1)A+\frac{g}{3}A\star A\star A\right) .  \label{action}
\end{equation}%
The kinetic term is given by the Virasoro operator which is a second order
differential operator acting on string fields $A\left( \bar{x},\xi ,\xi
^{gh}\right) $. We specify this operator later since it will be the main
focus of this section. The string field is a function of the midpoint
coordinate $\bar{x}$ and matter and fermionic $bc$ ghosts coordinates, $\xi
\equiv \left( x_{e}^{\mu },p_{e}^{\mu }\right) $, ($\mu =0,1,\cdots ,d-1,$ ($%
d=26$)) and $\xi ^{gh}\equiv \left(
x_{e}^{b},x_{e}^{c},p_{e}^{b},p_{e}^{c}\right) $. The even index $e$
specifies the mode and takes its value in $\left\{ e=2,4,\cdots ,2N\right\} $
where $N$ is the large integer which is introduced for the regularization. $%
\xi $ (resp. $\xi ^{gh}$) are Grassmann even (resp. odd) coordinates in
noncommutative Moyal space with the star product specified by \cite{B, BKM1,
PREP, Erler}\footnote{%
Compared to \cite{BKM1}, we redefined the Grassmann variables
with odd label to even
ones $x_{e}^{b},x_{e}^{c},p_{e}^{b},p_{e}^{c}$ as
\begin{equation}
x_{e}^{b}=\kappa _{e}^{-1}Sx_{o}^{gh}\,,\quad p_{e}^{b}=\kappa
_{e}Sp_{o}^{gh}\,,\quad x_{e}^{c}=Ty_{o}^{gh}\,,\quad p_{e}^{c}=\bar{R}%
q_{o}^{gh}\,,\quad (S:=\kappa _{e}T\kappa _{o}^{-1})\,.  \notag
\end{equation}%
The form of the Moyal $\star $ product (Eq.(70) in \cite{BKM1}) is invariant
under this transformation.}

\begin{equation}
\left[ x_{e}^{\mu },x_{e^{\prime }}^{\nu }\right] _{\star }=i\theta \delta
_{ee^{\prime }}\eta ^{\mu \nu }\,,\quad \left\{ x_{e}^{b},p_{e^{\prime
}}^{b}\right\} _{\star }=\left\{ x_{e}^{c},p_{e^{\prime }}^{c}\right\}
_{\star }={\theta ^{\prime }}\delta _{ee^{\prime }}\,\,.
\label{star_product}
\end{equation}%
In Appendix A we summarize MSFT notation that is used in this paper,
including precise definitions of the Bogoliubov transformation between
conventional open string operators and Moyal coordinates.

We note the additional free parameters in regulated MSFT. Namely, the
spectrum parameters $\kappa _{e}$ and $\kappa _{o}$ that correspond to
string oscillator frequencies can be chosen freely (see Appendix A for the
notation) as the definition of the regularized theory. While they will be
fixed at the end to $\kappa _{n}=n$ to reproduce the correct open string
defined in Eq.(\ref{open_string_limit}), the algebraic framework of MSFT is
well defined for any $\kappa _{n}$ as a function of $n.$

The equivalence of MSFT, with its regulator, to the conventional operator
formulation has already been established in the following sense
\cite{BKM1},

\begin{enumerate}
\item The spectrum of the kinetic term is identical to the conventional open
string at large $N$. That is, the propagator in perturbation theory is
identical in both theories.

\item The Neumann coefficients of the oscillator formalism were computed
directly from the Moyal product \cite{BM2}\cite{PREP}, and these were shown
to satisfy the Gross-Jevicki nonlinear relations \cite{GJ} for any
frequencies $\kappa _{n}$. Their simple expression explains as well agrees
with spectroscopy \cite{spectroscopy}, and agree numerically with other
computations at large $N$.
\end{enumerate}

These facts are sufficient to guarantee the equivalence of two formulations
of string field theory in the computation of any perturbative string
amplitudes \cite{BKM1}. Our formulation is well adapted for the discussion
of nonperturbative string physics which will be the main topic of this paper.

We note that the conformal symmetry is lost in the regularized theory since
we truncate the number of oscillators at level $2N$. Such truncation is,
however, indispensable for the correct treatment of the associativity
anomaly in the fundamental matrices \cite{BM1} and also the Neumann
coefficients. We note that every explicit computation of string field theory
so far is based on some cut-off (level-truncation and so on). The additional
freedom in the spectrum is actually directly related to the loss of
conformal symmetry. We have to fix it at the end of the computation by
taking the large $N$ limit.

One of the main goals of string field theory (and also the main goal of this
paper) is to solve the nonlinear equation of motion,
\begin{equation}
(L_{0}-1)A+{\alpha ^{\prime }}gA\star A=0\,\,.  \label{eom}
\end{equation}%
The solutions, except for the trivial one ($A=0$), describe nonperturbative
backgrounds of open string theory which should be related to D-branes. It is
widely believed that there exists a unique solution $A_{0}$ which describes
the tachyon vacuum where D-branes are annihilated. There exist a large
amount of numerical evidence \cite{Numerics} that confirms the existence and
the desirable properties of such solution (D-brane tension, the absence of
open string propagation and so on).

While the achievement of the numerical study is very impressive, it is still
indispensable to develop an analytic framework to study the tachyon vacuum
and other nonperturbative phenomena in string theory. In the operator
formulation a difficulty originates from the fact that the basis which
diagonalizes the kinetic term (conventional Fock space representation, which
is its main tool) gives a complicated expression for the star product
including the Neumann coefficients.

In the Moyal star formulation, on the other hand, the star product that
produces (\ref{star_product}) is trivial, while the kinetic term becomes
off-diagonal. The complication of the kinetic term is, however, manageable
and has not been a hindrance in developing the formalism for practical
computations, such as Feynman graphs \cite{BKM1}, and as we will see in the
following, for nonperturbative solutions.

\subsection{Splitting of kinetic term}
\label{sub:splitting}

We first translate the Virasoro operator from the conventional operator
language to the Moyal star formulation by following the path in references
\cite{BM2,BKM1,PREP}.

The Virasoro operator $L_{0}$ written in terms of the standard oscillator
notation is
\begin{equation}
L_{0}^{osc}=\frac{1}{2}\alpha _{0}^{2}+\sum_{n=1}^{2N}\eta _{\mu \nu }\alpha
_{-n}^{\mu }\alpha _{n}^{\nu }+\sum_{n=1}^{2N}\kappa
_{n}(b_{-n}c_{n}+c_{-n}b_{n})
\end{equation}%
where we explicitly truncate the number of oscillators and rewrite the
frequency from $n$ to $\kappa _{n}$. The commutation relation among
oscillators with generic frequency is given in Eq.(\ref{osc_comm}). After
translating the oscillators into the Moyal space as in Appendix A, we arrive
at the expression of $L_{0}$ as a second order differential operator acting
on the Moyal field $A$ in the Siegel gauge ($\theta ,{\theta ^{\prime }}$
are arbitrary parameters which define the noncommutativity, and $l_{s}=\sqrt{%
2{\alpha ^{\prime }}}$),
\begin{eqnarray}
L_{0} &=&L_{0}^{matter}+L_{0}^{ghost}\,\,, \\
L_{0}^{matter} &=&\sum_{e>0}\left( -\frac{l_{s}^{2}}{2}\frac{\partial ^{2}}{%
\partial x_{e}^{2}}-\frac{\theta ^{2}}{8l_{s}^{2}}\kappa _{e}^{2}\frac{%
\partial ^{2}}{\partial p_{e}^{2}}+\frac{1}{2l_{s}^{2}}\kappa
_{e}^{2}x_{e}^{2}+\frac{2l_{s}^{2}}{\theta ^{2}}p_{e}^{2}\right) +\frac{1}{2}%
\left( 1+\bar{w}w\right) \beta _{0}^{2}  \notag \\
&&+\frac{il_{s}}{2}\beta _{0}\sum_{e>0}w_{e}\frac{\partial }{\partial x_{e}}-%
\frac{1}{1+\bar{w}w}\frac{2l_{s}^{2}}{\theta ^{2}}\left(
\sum_{e>0}w_{e}p_{e}\right) ^{2}-\frac{d}{2}\sum_{n=1}^{2N}\kappa _{n}
\,\,, \\
L_{0}^{ghost} &=&i\sum_{e>0}\left( \frac{\partial }{\partial x_{e}^{b}}\frac{%
\partial }{\partial x_{e}^{c}}+\frac{{\theta ^{\prime }}^{2}}{4}\kappa
_{e}^{2}\frac{\partial }{\partial p_{e}^{b}}\frac{\partial }{\partial
p_{e}^{c}}+\kappa _{e}^{2}x_{e}^{b}x_{e}^{c}+\frac{4}{{\theta ^{\prime }}^{2}%
}p_{e}^{b}p_{e}^{c}\right)   \notag \\
&&-\frac{i}{1+\bar{w}w}\left( \sum_{e}w_{e}\frac{\partial }{\partial
x_{e}^{b}}\right) \left( \sum_{e^{\prime }}w_{e^{\prime }}\frac{\partial }{%
\partial x_{e^{\prime }}^{c}}\right) +\sum_{n=1}^{2N}\kappa _{n}\,\,.
\end{eqnarray}%
$\beta _{0}=-il_{s}\frac{\partial }{\partial \bar{x}}=-il_{s}\frac{\partial
}{\partial x_{0}}$ represents the center of mass momentum, which can be
written as a derivative of the midpoint coordinate $\bar{x}$. We note that
the expression consists of mostly diagonal combination of the Moyal
variables. The off-diagonal pieces with coefficient $(1+\bar{w}w)^{-1}$
appear because of the Bogoliubov transformation from the odd modes with
spectrum $\kappa _{o},$ to the even modes with spectrum $\kappa _{e}$. This
complication of the kinetic term is the cost of the simplification of the
star product in the Moyal product formulation.

At this stage, the equation of motion appears as horribly
complicated -- it is nonlinear, it contains an infinite order of
derivatives through the star product, and is off-diagonal through
the terms proportional to $(1+\bar{w}w)^{-1}$. However, there
exists a critical simplification of $L_{0}$ which saves us from
most of these difficulties. The trick is to rewrite the diagonal
pieces of $L_{0}$ by using the star product in the following form.
It gives the splitting of $L_{0}$ into two parts containing the symbols $%
\mathcal{L}_{0}$,$\gamma $ \cite{BM2,BKM1,PREP},
\begin{equation}
(L_{0}-1)A=\left( \mathcal{L}_{0}(\beta _{0})\star A+A\star \mathcal{L}%
_{0}(-\beta _{0})\right) +\gamma A,
\end{equation}%
with \footnote{%
The expression $A\star {\mathcal{L}_{0}}(-\beta _{0})$ is to be understood
that $\beta _{0}$ is a derivative applied on $A(\bar{x})$ on its left, even
though it is written on the right for convenience of notation.}
\begin{eqnarray}
\mathcal{L}_{0}(\beta _{0}) &=&\mathcal{L}_{0}^{matter}(\beta _{0})+\mathcal{%
L}_{0}^{ghost}-\frac{1}{2}\,\,,\qquad \gamma =\gamma ^{matter}+\gamma
^{ghost} \,\,,\\
\mathcal{L}_{0}^{matter}(\beta _{0}) &=&\sum_{e>0}\left( {\frac{l_{s}^{2}}{%
\theta ^{2}}}p_{e}^{2}+{\frac{\kappa _{e}^{2}}{4l_{s}^{2}}}x_{e}^{2}-{\frac{%
l_{s}}{\theta }}w_{e}p_{e}\beta _{0}\right) +{\frac{1}{4}}(1+\bar{w}w)\beta
_{0}^{2}-{\frac{d}{4}}\sum_{n>0}\kappa _{n}\,\,,  \notag \\
{\mathcal{L}_{0}}^{ghost} &=&i\sum_{e>0}\left( \frac{2}{{\theta ^{\prime }}%
^{2}}p_{e}^{b}p_{e}^{c}+\frac{\kappa _{e}^{2}}{2}x_{e}^{b}x_{e}^{c}\right) +%
\frac{1}{2}\sum_{n>0}\kappa _{n}\,,  \label{L0} \\
\gamma ^{matter} &=&-{\frac{1}{1+\bar{w}w}}{\frac{2l_{s}^{2}}{\theta ^{2}}}%
\left( \sum_{e>0}w_{e}p_{e}\right) ^{2}\,\,,  \notag \\
\gamma ^{ghost} &=&-\frac{i}{1+\bar{w}w}\left( \sum_{e>0}{w}_{e}\frac{%
\partial }{\partial x_{e}^{b}}\right) \left( \sum_{e^{\prime }>0}{w}%
_{e^{\prime }}\frac{\partial }{\partial x_{e^{\prime }}^{c}}\right) \,\,.
\label{gamma}
\end{eqnarray}%
The star products with the \textit{field} ${\mathcal{L}_{0}}$ reproduces the
diagonal part of the \textit{differential operator} $L_{0}-1,$ while the
ordinary product with $\gamma $ gives the off-diagonal part of $L_{0}-1$.

The action of $\gamma $ can also be written by using the star product while
it cannot be split as left and right multiplications alone,
\begin{eqnarray}
\gamma ^{matter}A &=&-\frac{l_{s}^{2}}{2\theta ^{2}(1+\bar{w}w)}%
\sum_{e,e^{\prime }}w_{e}w_{e^{\prime }}\eta _{\mu \nu }\left\{ p_{e}^{\mu
},\left\{ p_{e^{\prime }}^{\nu },A\right\} _{\star }\right\} _{\star }\,\,,
\\
\gamma ^{ghost}A &=&-\frac{i}{{\theta ^{\prime }}^{2}(1+\bar{w}w)}%
\sum_{e,e^{\prime }}w_{e}w_{e^{\prime }}\left\{ p_{e}^{b},\left[
p_{e^{\prime }}^{c},A\right] _{\star }\right\} _{\star }\,\,.
\end{eqnarray}%
These formulae will be useful in the concrete computation in section \ref%
{perturbation}.

We emphasize that $\gamma $ depends only on the rank one quantity $\hat{p}%
^{\mu ,b,c}=\left( 1+\bar{w}w\right) ^{-1/2}\sum_{e}w_{e}p_{e}^{\mu ,b,c}$,
which is basically a single string momentum mode. This mode was first
pointed out as the source of associativity anomalies of the star product in
string field theory \cite{BM1}. The strange mode that was later discovered
in the continuous Moyal formalism \cite{DLMZ} is given by $\bar{p}^{\mu
,b,c}=\left( 1+\bar{w}w\right) ^{-1/2}\hat{p}^{\mu ,b,c}$ up to a numerical
constant as shown in \cite{BM2}\cite{B2}. As we will see, it is due to this
mid-point complication that string field theory is not trivially solvable.
We describe our strategy to solve the equation of motion with this splitting
in the next subsection.

We explain the basic role of the two parts of $L_{0}$. If the $\gamma $ term
were absent, the spectrum of the free string would depend only on the
frequencies $\kappa _{e}$ since ${\mathcal{L}_{0}}$ (aside from the last
constant term) depends only on $\kappa _{e}.$ The string spectrum coming
only from the two terms with ${\mathcal{L}_{0}}$ looks like the string
spectrum of both even and odd oscillators, but with the frequency of the odd
oscillator $\kappa _{o}$ adjusted to be equal to the frequency of the
neighboring even oscillator $\kappa _{o}\rightarrow \kappa _{o+1}=\kappa
_{e}.$ On further reflection, one can see that the ${\mathcal{L}_{0}}$ term
by itself essentially describes the kinetic term of the two half strings
which the original open string is composed of. However, the open string has
a different spectrum, namely the even modes have frequencies $\kappa _{e}$
and the odd modes have frequencies $\kappa _{o}$. The second term $\gamma $
precisely fixes the discrepancy of the spectrum from the half string
description. It carries the information of the midpoint where two half
strings are connected, as shown in \cite{BM2}\cite{BKM1}. In this sense, we
will refer to $\gamma $ as the \textquotedblleft midpoint
correction\textquotedblright .

The factor $(1+\bar{w}w)^{-1}$ appears to vanish naively in the open string
limit Eq.(\ref{open_string_limit}). However, this is misleading because in
computations one obtains factors of $\bar{w}w$ in the numerator that produce
finite contributions by the $\gamma $ term. This is the mechanism of
anomalies \cite{BM1}. It is because of this subtlety that this term has been
largely missed in the split string literature.

We note that the splitting of $L_{0}$ into ${\mathcal{L}_{0}}$ part and $%
\gamma $ term is not unique. Namely one may obtain the same $L_{0}$ by the
shift,
\begin{equation}
{\mathcal{L}_{0}}\rightarrow {\mathcal{L}_{0}}-f(\xi ,\xi ^{gh})\,,\quad
\gamma A\rightarrow \gamma A+\left\{ f(\xi ,\xi ^{gh}),A\right\} _{\star
}\,\,.
\end{equation}%
The framework of our analysis explained in the next subsection will not be
basically affected by such a change as long as $f(\xi )$ is quadratic with
respect to $\xi ,\xi ^{gh}$. It modifies, however, the half string spectrum
and the Fock space structure discussed in the following sections. A fact
mentioned above, namely that $\gamma $ depends only on the rank one quantity
$\hat{p}$, will be generally broken in such an arbitrary shift. Such a
complicated choice of the splitting between ${\mathcal{L}_{0}}$ and $\gamma $
may be useful to define a sensible perturbation expansion in the open string
limit of Eq.(\ref{open_string_limit}). We will come back to this issue later.

\subsection{Strategy}

Due to the separation of the kinetic term, we rewrite the action in the
following form,
\begin{eqnarray}
S &=&-\int d^{d}\bar{x}\,Tr\left( \frac{1}{2{\alpha ^{\prime }}}A\star
(L_{0}-1)A+\frac{g}{3}A\star A\star A\right) =-\left(
S_{1}+S_{2}+S_{3}\right) \,\,. \\
S_{1} &=&\frac{1}{{\alpha ^{\prime }}}\int d^{d}\bar{x}\,TrA\star {\mathcal{L%
}_{0}}(\beta _{0})\star A\,\,, \\
S_{2} &=&\frac{1}{2{\alpha ^{\prime }}}\int d^{d}\bar{x}\,TrA\star (\gamma
A)\,\,, \\
S_{3} &=&\frac{g}{3}\int d^{d}\bar{x}\,TrA\star A\star A\,\,.
\end{eqnarray}

Before pursuing the nonperturbative analysis, let us emphasize that the
conventional perturbation expansion of open string field theory is
successfully reproduced in our MSFT formalism. In the conventional
perturbative case, the classical equation of motion that comes from the
quadratic part $S_{1}+S_{2}$, namely $\left( L_{0}-1\right) A=0,$ gives the
spectrum of string states associated with the oscillator frequencies $%
(\kappa _{e},\kappa _{o})$ \cite{BM2}\cite{BKM1}. The interaction among
these perturbative states is given by $S_{3}$ where only the use of the
Moyal product is sufficient to compute interactions. Pursuing this in our
formalism gives results that are in agreement with the conventional
oscillator approach to the open string field theory \cite{GJ}. In
particular, our MSFT formalism including ghosts, replaces the complicated
Neumann coefficients of the oscillator approach to express the interaction
part explicitly \cite{BM2}\cite{PREP}. Our theory includes a consistent
regulator, and with it numerical estimates of certain quantities have been
compared successfully to numerical results obtained in other formalisms \cite%
{PREP}. Thus, we are confident that we have the correct theory to explore
nonperturbative phenomena.

By dividing the kinetic term into $S_{1}$ and $S_{2}$, we can pursue the
alternative splitting of the action. Namely we first solve the system with $%
S_{1}+S_{3}$. As we discuss in the following sections, there is a basis
which diagonalizes these two terms (${\mathcal{L}_{0}}$ and $\star $%
-product) at the same time\footnote{%
There is another basis which diagonalize $S_{2}$ and $S_{3}$ at the same
time and we may carry out the program which parallels our discussion in the
following (see appendix \ref{s:particle_limit}). However, there is no basis
which diagonalized all three terms $S_{i}$ ($i=1,2,3$). It gives the
essential difficulty to obtain the tachyon vacuum in the analytic form.}. It
gives a major simplification of the equation of motion and we can solve the
nonlinear equation analytically at the classical level. The solutions are
given in terms of the projection operators and should be regarded as
defining nonperturbative vacua of open string field theory in the limit $%
S_{2}\rightarrow 0$. This is in some sense similar to the VSFT proposal \cite%
{VSFT} although we expand the system from the different vacuum. The midpoint
correction $S_{2}$ will be introduced as the \textquotedblleft
perturbation\textquotedblright\ to the exact solutions of $S_{1}+S_{3}.$

At the level of the equation of motion, this strategy is equivalent to
writing (\ref{eom}) as\footnote{%
We omit the $\bar{x}$ derivative (or $\beta _{0}$) since our main focus in
this paper is the study of the translational invariant solutions.},
\begin{equation}
{\mathcal{L}_{0}}\star A+A\star {\mathcal{L}_{0}}+{\alpha ^{\prime }}gA\star
A=-\epsilon \gamma A\,\,.  \label{eom2}
\end{equation}%
and treat the right hand side as a source term. We introduced formally an
expansion parameter $\epsilon $ which will be used to describe the order of
perturbation. We must put $\epsilon =1$ at the end. We will first solve the
equation in the absence of the source term, and later include the source for
the complete solution.

At first glance, even without the source term, we are still left with a
nonlinear differential equation of infinite order and the analytic study of
such an equation of motion seems impossible. However, here the methods of
noncommutative geometry come in handy for any operator ${\mathcal{L}_{0}.}$
Thus, at the formal level, one may find solutions $A=A_{P}$ labeled by
projectors $P$ in the following form,
\begin{equation}\label{e-Ap}
A_{P}=-\frac{2}{{\alpha ^{\prime }}g}{\mathcal{L}_{0}}\star P.
\end{equation}%
Here $P$ is any projector which satisfies the following properties,
\begin{equation}
P\star P=P\,\,,\qquad \left[ P,{\mathcal{L}_{0}}\right] =0\,\,.
\end{equation}%
Once we have a solution $A_{P}$ of the homogeneous equation$,$ the
corrections to it due to $\gamma $ are taken into account by the following
integral equation which is equivalent to an exact formal solution\footnote{%
To verify this, consider the star anticommutator of both sides of the
integral equation with the quantity $\left( \mathcal{L}_{0}+\alpha ^{\prime
}gA_{P}\right) .$ The left side is $\left\{ \left( \mathcal{L}_{0}+\alpha
^{\prime }gA_{P}\right) ,A\right\} _{\star }$ while the right side, in
addition to $\left\{ \left( \mathcal{L}_{0}+\alpha ^{\prime }gA_{P}\right)
,A_{P}\right\} _{\star }$ , produces a total $\tau $ derivative under the
integral sign%
\begin{equation}
\int_{0}^{\infty }d\tau \;\frac{\partial }{\partial \tau }\left[ e_{\star
}^{-\tau \left( \mathcal{L}_{0}+\alpha ^{\prime }gA_{P}\right) }\star \left(
\alpha ^{\prime }g\left( A-A_{P}\right) _{\star }^{2}+\epsilon
\gamma A\right) \star
e_{\star }^{-\tau \left( \mathcal{L}_{0}+\alpha ^{\prime }gA_{P}\right) }%
\right] .
\end{equation}%
Assuming a positive spectrum for the hamiltonian $\left( \mathcal{L}%
_{0}+\alpha ^{\prime }gA_{P}\right) ,$ the integral contributes only at the
boundary $\tau =0.$ We will return later to discuss the issue of the
spectrum of $\left( \mathcal{L}_{0}+\alpha ^{\prime }gA_{P}\right) ,$ for
now we proceed formally. Inserting the result of the integral, the left and
right sides of the equation yield%
\begin{equation}
\left\{ \left( \mathcal{L}_{0}+\alpha ^{\prime }gA_{P}\right)
,A\right\} _{\star }=\left\{ \left( \mathcal{L}_{0}+\alpha
^{\prime }gA_{P}\right) ,A_{P}\right\} _{\star }- \left( \alpha
^{\prime }g\left( A-A_{P}\right) _{\star }^{2}+\epsilon\gamma
A\right) .
\end{equation}%
Rearranging this equation we obtain back our full equation of motion in Eq.(%
\ref{eom2}) after using the fact that $A_{P}$ is a solution of the
homogeneous equation.\label{intEq}} of Eq.(\ref{eom2})%
\begin{equation}
A=A_{P}-\int_{0}^{\infty }d\tau \;e_{\star }^{-\tau \left( \mathcal{L}%
_{0}+\alpha ^{\prime }gA_{P}\right) }\star \left( \alpha ^{\prime }g\left(
A-A_{P}\right) _{\star }^{2}+\epsilon \gamma A\right) \star e_{\star
}^{-\tau \left( \mathcal{L}_{0}+\alpha ^{\prime }gA_{P}\right) }.
\label{int}
\end{equation}%
Since $A$ appears on both sides, one approach to obtaining an explicit
solution for $A$ is by recursion. As will be discussed below, this amounts
to a perturbative expansion of $A$ in powers of $\epsilon ,$
\begin{equation}
A=A^{\left( 0\right) }+\epsilon A^{\left( 1\right) }+\epsilon ^{2}A^{\left(
2\right) }+\cdots \,\,,
\end{equation}%
with the lowest order being $A^{\left( 0\right) }=A_{P}$. The full
perturbative series is given later in section \ref{perturbation}. This
analysis could in fact be pursued for any ${\mathcal{L}_{0}.}$

A natural question is (1) does such projector $P$ indeed exist, (2) can all
such projectors be written explicitly for given ${\mathcal{L}_{0}}$, (3)
does this exhaust all the solutions of Eq.(\ref{eom2}) with $\epsilon =0$?
The answer is formally yes to all three questions, as follows.

As in the situation in the noncommutative soliton \cite{GMS}, the oscillator
representation of the Moyal product gives an equivalent but more transparent
means to analyze such a problem. In this language, we
take $\mathcal{L}_{0}$ as a hamiltonian.
The rank one projectors which commute with $\cL$
can be constructed schematically
as outer products $P_{\lambda }=|\lambda \rangle
\langle\lambda |$
of the normalized eigenstates of the hamiltonian
$\mathcal{L}_{0}|\lambda\rangle=\lambda |\lambda \rangle$, with
$\langle\lambda |\lambda ^{\prime }\rangle=\delta _{\lambda
\lambda ^{\prime }}.$ Finding solutions of the form (\ref{e-Ap})
reduces to finding eigenstates of $\mathcal{L}_{0}$. But
this is an easy task for our $\mathcal{L}_{0}$ since it is the hamiltonian
of a collection of harmonic oscillators\footnote{%
 In the Moyal language, the
 projectors we want as functions of the noncommutative space $P_{\lambda
 }\left( \xi \right) $ are known as the Wigner distributions for all the
 quantum states of the harmonic oscillator. These are well known in the
 literature in the case of a single harmonic oscillator \cite{wigner}.}.
A careful treatment along this scenario is given in the next section.

After we find an analytic form for $A^{\left( 0\right) }$, we use Eq.(\ref%
{eom2}) recursively to determine the expansion of the analytic solution of
full equation of motion,
\begin{eqnarray}
\left\{ {\mathcal{L}_{0}}^{\prime },A^{\left( k\right) }\right\} _{\star }
&=&-\gamma A^{\left( k-1\right) }-{\alpha ^{\prime }}g\sum_{i=1}^{k-1}A^{%
\left( i\right) }\star A^{\left( k-i\right) }\,\,,  \label{recursion} \\
{\mathcal{L}_{0}}^{\prime } &\equiv &{\mathcal{L}_{0}}+{\alpha ^{\prime }}%
gA^{\left( 0\right) }\,\,.
\end{eqnarray}%
This is, of course, equivalent to the iterative solution of the integral
equation in Eq.(\ref{int}), but we will use a matrix formalism that is
convenient in the case of oscillators. We will show that there is no
obstruction to solving Eq.(\ref{recursion}) order by order and one can
determine $A^{\left( k\right) }$ uniquely for any starting point $A^{\left(
0\right) }$. The explicit form of the first order correction and formal
solution for any $A^{\left( k\right) }$ is given in section \ref%
{perturbation}.

\section{Physics at $\protect\gamma=0$}

\subsection{Splitting limit}

The nature of the system at zeroth order in the $\gamma $ term may be
understood as follows. Let us begin by examining $L_{0}$ in the absence of
the gamma term.
The remaining part of $L_{0}$ has no information about the odd frequencies $%
\kappa _{o}$ and the resulting spectrum of the modified $L_{0}$ corresponds
to string oscillators that are both even and odd, but with the odd
frequencies $\kappa _{o}$, instead of being arbitrary, replaced by the
neighboring even frequency $\kappa _{e}$%
\begin{equation}
\kappa _{o}\rightarrow \kappa _{o+1}=\kappa _{e}\,\,,\;\mbox{for  }%
\,\,o=1,3,5,\cdots ,  \label{splitting}
\end{equation}%
while $\kappa _{e}$ is still arbitrary as is usual in regulated MSFT.
This exactly characterizes the spectrum produced by the $\mathcal{L}_{0}$
part of $L_{0}$ in Eq.(\ref{eom2}) in the absence of $\gamma $. Indeed, an
inspection of $\mathcal{L}_{0}$ shows that it contains only the even
frequencies $\kappa _{e}.$ Given the fact that the star product is
independent of the frequencies, the system $S_{1}+S_{3}$ has no information
on how to correct the frequencies of the odd oscillators from $\kappa _{o+1}$
to $\kappa _{o}.$

Conversely if we insert Eq.(\ref{splitting}) in the regulated MSFT system, a
major simplification occurs in the defining matrices and vectors $U,v,w$.
Namely $U$ becomes the identity matrix and $v,w$ vanishes. The vanishing of $%
w_{e}$ immediately implies $S_{2}=0$, and the open string splits into
independent half strings. We recall that the meaning of $U,v,w$ is to give
Bogoliubov transformation from variables with frequency $\kappa_o$ to those
with $\kappa_e$. If there is no difference between the sets of frequencies,
there is no need to perform Bogoliubov transformation and the matrix which
define it becomes trivial.

The MSFT formalism gives us the ability to consider this limit as well as
the corrections. Intuitively one may think that a small change in the
frequencies may not change the physics of the system drastically. Indeed,
Eq.(\ref{splitting}) is a small change in $\kappa _{o}$ when $o$ is
sufficiently large. Hence a more interesting situation is to consider MSFT
in the limit (\ref{splitting}) for $o$ larger than some number, $o>2N$ while
leaving both $\kappa _{e},\kappa _{o}$ arbitrary for $e,o\leq 2N.$ In that
case the trivialization of $U,v,w$ applies only to the modes above $2N,$ and
$\gamma $ gets contributions only from the modes up to $2N.$ As long as $%
\gamma \neq 0$ the string does not split into two independent halves. So, it
seems worth studying such limits, at least for the higher modes, since the
formalism simplifies drastically while the physics (depending on the
specific question) may be about the same. Furthermore, since we have
complete control of the corrections, one can test both analytically and
numerically the size of the correction. Such more complicated approximation
scenarios are under study. But in the present paper we do not take any
limits on the $\kappa _{e},\kappa _{o};$ we simply study the system in
powers of $\gamma $ but for arbitrary $\kappa _{e},\kappa _{o}.$

\subsection{Oscillator representation}

\label{s:matrix} 
The simplification achieved in the splitting limit is elegantly rewritten in
the oscillator notation\footnote{%
We note that a somewhat similar representation was considered in \cite%
{Okuyama_et_al} in the expansion around the sliver solution by neglecting
the midpoint correction.}. We diagonalize the action of ${\mathcal{L}_{0}}$
and the star product at the same time in the matrix notation.

We introduce the creation and annihilation operators of the matter sector
\cite{BM2} and the ghost sector \cite{BKM1} \cite{PREP} (in even mode
variable)
\begin{eqnarray}
&&\beta _{e}^{\mu }=\frac{1}{\sqrt{\kappa _{e}}}\left( -i\epsilon (e)\frac{%
\kappa _{e}}{2l_{s}}x_{e}^{\mu }+\frac{l_{s}}{\theta }p_{e}^{\mu }\right) ,\,
\label{beta-osc} \\
&&\beta _{e}^{b}=\left( -i\epsilon (e)\frac{{\kappa }_{e}}{2}x_{e}^{b}+\frac{%
1}{{\theta ^{\prime }}}p_{e}^{c}\right) ,\,\quad \beta _{e}^{c}=\left( \frac{%
1}{2}x_{e}^{c}-\epsilon (e)\frac{i}{{\theta ^{\prime }\kappa }_{e}}%
p_{e}^{b}\right)  \label{beta-osc-3}
\end{eqnarray}%
which satisfy the canonical commutation relations with respect to $\star $
product,
\begin{equation}
\left[ \beta _{e}^{\mu },\beta _{e^{\prime }}^{\nu }\right] _{\star }=\eta
^{\mu \nu }\epsilon(e)
\delta _{e+e^{\prime }}\,,\qquad \left\{ \beta _{e}^{b},\beta
_{e^{\prime }}^{c}\right\} _{\star }=\delta _{e+e^{\prime }}\,.
\label{osc_star}
\end{equation}%
In terms of these oscillators, one can rewrite ${\mathcal{L}_{0}}$ as
\begin{equation}
{\mathcal{L}_{0}}=\sum_{e>0}\kappa _{e}\left( \beta _{-e}^{\mu }\star \beta
_{e}^{\nu }\eta _{\mu \nu }+\beta _{-e}^{b}\star \beta _{e}^{c}+\beta
_{-e}^{c}\star \beta _{e}^{b}\right) -\nu \,,\quad \nu =\frac{1}{2}-\frac{d-2%
}{4}(\sum_{e>0}\kappa _{e}-\sum_{o>0}\kappa _{o})\,\,,  \label{nu_and_cL}
\end{equation}%
We note that both the star product (\ref{osc_star}) and the kinetic term (%
\ref{nu_and_cL}) are diagonal with respect to the basis (\ref{beta-osc}--\ref%
{beta-osc-3}). It makes the splitting limit completely solvable.

We introduce the (nonperturbative) vacuum state through the relations,
\begin{equation}
\beta _{e}^{\mu }\star A_{0}=\beta _{e}^{b}\star A_{0}=\beta _{e}^{c}\star
A_{0}=A_{0}\star \beta _{-e}^{\mu }=A_{0}\star \beta _{-e}^{b}=A_{0}\star
\beta _{-e}^{c}=0\,\,.
\end{equation}%
The state that solves them becomes
\begin{equation}
A_{0}\sim \exp \left( -\sum_{e}\left( \frac{\kappa _{e}}{2l_{s}^{2}}%
(x_{e})^{2}+\frac{2l_{s}^{2}}{\theta ^{2}\kappa _{e}}(p_{e})^{2}+i\kappa
_{e}x_{e}^{b}x_{e}^{c}+\frac{4i}{{\theta ^{\prime }}^{2}\kappa _{e}}%
p_{e}^{b}p_{e}^{c}\right) \right) \,.  \label{butterfly}
\end{equation}%
This is called \textquotedblleft butterfly state" \footnote{%
In fact, we can show that this $A_{0}$ corresponds to (twisted) butterfly
state: $e^{-{\frac{1}{2}}(L_{-2}^{m}+L_{-2}^{^{\prime }g})}c_{1}|0\rangle $
in the limit $\kappa _{e}=e,\kappa _{o}=o,N=\infty $ \cite{PREP}.} in the
literature \cite{VSFT} and satisfies the projector condition,
\begin{equation}
A_{0}\star A_{0}=A_{0}\,,\quad A_{0}^{\ast }=A_{0}\,\,.
\end{equation}%
This is the critical simplification by neglecting the mixing term. As we see
in the following, the states generated by multiplying the creation operators
from the left or the annihilation operators from the right diagonalize both $%
{\mathcal{L}_{0}}$ and the star product.

To illustrate our idea we use a simplified situation with only one pair
oscillators $a$ and $a^{\dagger }$ satisfying $[a,a^{\dagger }]=1$. The
orthonormal basis is given explicitly as,
\begin{eqnarray}
&&\phi _{nm}(x,p)=\frac{1}{\sqrt{n!m!}}(a^{\dagger })_{\star }^{n}\star
A_{0}\star (a)_{\star }^{n}\,,\quad \phi _{nm}\star \phi _{rs}=\delta
_{mr}\phi _{ns}\,\,, \\
&&\hat{N}\star \phi _{nm}=n\phi _{nm}\,,\quad \phi _{nm}\star \hat{N}=m\phi
_{nm}\,,\quad \hat{N}=a^{\dagger }a\,\,.
\end{eqnarray}%
The orthogonality with respect to $\star $ product is essential in the
following. In the above case, it can be proved by the commutation relation
and the condition for $A_{0}$ (\ref{butterfly}). $\phi _{nn}$ becomes
mutually orthogonal projectors,
\begin{equation}
\phi _{nn}\star \phi _{mm}=\delta _{nm}\phi _{mm}\,\,.
\end{equation}

The multi-oscillator extension of above basis is given simply by direct
product. We introduce the multi-index symbol
\begin{equation}
\mathbf{n}=\left\{ n_{e}^{i}|n_{e}^{i}\geq 0,e=2,4,\cdots ,2N,\;i=0,\cdots
,d;\,\,n_{e}^{i}=0,1\mbox{ for }\;i=b,c\right\} \,\,,
\end{equation}%
and introduce the states $\phi _{\mathbf{n}\mathbf{m}}=\prod_{e}\prod_{i}%
\phi _{n_{e}^{i},m_{e}^{i}}^{\left\{ i,e\right\} }$. We denote the set of
multi-indices $\left\{ \mathbf{n}\right\} =\mathbf{Z}^{\otimes Nd}\otimes
\mathbf{Z}_{2}^{\otimes 2N}$ as $\mathcal{B}$. The basis satisfies
\begin{eqnarray}
&&{\mathcal{L}_{0}}\star \phi _{\mathbf{n}\mathbf{m}}=\lambda _{\mathbf{n}%
}\phi _{\mathbf{n}\mathbf{m}}\,,\quad \phi _{\mathbf{n}\mathbf{m}}\star {%
\mathcal{L}_{0}}=\lambda _{\mathbf{m}}\phi _{\mathbf{n}\mathbf{m}}\,,\quad
\lambda _{\mathbf{n}}\equiv \left( \sum_{i}\sum_{e>0}\kappa
_{e}n_{e}^{i}\right) -\nu \,\,, \\
&&\phi _{\mathbf{n}\mathbf{m}}\star \phi _{\mathbf{r}\mathbf{s}}=\delta _{%
\mathbf{mr}}\phi _{\mathbf{n}\mathbf{s}}\,\,.
\end{eqnarray}

We expand $A=\sum_{\mathbf{nm}}a_{\mathbf{nm}}(\bar{x})\phi _{\mathbf{mn}}$
and put it in the Lagrangian, we obtain $S_{1}+S_{3}=S^{matrix}+\delta S$
with 
\begin{eqnarray}
S^{matrix} &=&\int d^{d}\bar{x}\,\mbox{Tr}\left( \frac{1}{2}\partial _{\bar{x%
}}a(\bar{x})\cdot \partial _{\bar{x}}a(\bar{x})+\frac{1}{{\alpha ^{\prime }}}%
\Lambda \cdot a\cdot a+\frac{g}{3}a\cdot a\cdot a\right) \,\,,
\label{matrix_model} \\
\delta S &=&\frac{\bar{w}w}{2}\int d^{d}\bar{x}\,\mbox{Tr}\,\partial _{\bar{x%
}}a\cdot \partial _{\bar{x}}a+\frac{2i}{\theta }\int d^{d}\bar{x}\,\mbox{Tr}%
\,a\cdot (\sum_{e}w_{e}P_{e})\cdot \partial _{\bar{x}}a\,\,.
\label{matrix_corr}
\end{eqnarray}%
where $\Lambda _{\mathbf{nm}}=\lambda _{\mathbf{n}}\delta _{\mathbf{nm}}$,
the trace $\mbox{Tr}$ is over the $\mathbf{n}$ indices and $\cdot $ is the
matrix product. $P_{e}$ is the matrix that corresponds to the star
multiplication of $p_{e}=\frac{\sqrt{\kappa _{e}}\theta }{2l_{s}}(\beta
_{e}+\beta _{e}^{\dagger })$. $\delta S$ becomes off-diagonal but does not
affect the equation of motion with the translational invariance. In the
splitting limit (\ref{splitting}), $\delta S$ vanishes because $w=0$. $%
S^{matrix}$ has a structure which is very similar to $c=d$ matrix model
except that the kinetic term contains a mass term $\Lambda $ which is not
proportional to the identity matrix. It describes a characteristic feature
of string theory that the color index $\mathbf{n}$ has a certain mass $%
\lambda _{\mathbf{n}}$.

\subsection{Translational invariant solutions}

In this section, we construct the translational invariant (=independent of $%
\bar{x}$) solutions in the splitting limit. It is quite interesting that the
equation of motion becomes completely solvable and we can give an explicit
form of the arbitrary solutions. Each solution describes a nonperturbative
vacuum of string field theory while it may be stable or unstable. We also
derive the open string spectrum around each vacuum explicitly while it
becomes rather trivial.

The equation of motion obtained from the action (\ref{matrix_model},\ref%
{matrix_corr}) is
\begin{equation}
(\lambda _{\mathbf{n}}+\lambda _{\mathbf{m}})a_{\mathbf{nm}}+{\alpha
^{\prime }}g\sum_{\mathbf{k}}a_{\mathbf{nk}}a_{\mathbf{km}}=0\,\,.
\end{equation}%
This equation has the following significant property which we call
\textquotedblleft reducibility\textquotedblright . Namely for any (finite)
subset of $\mathcal{B}^{\prime }=\left\{ \mathbf{k}_{1},\cdots ,\mathbf{k}%
_{n}\right\} \in \mathcal{B}$, one can consistently restrict the equation of
motion by replacing $a$ to  its rank $n$ sub-matrix $a_{\mathbf{kl}}$
with $\mathbf{k},\mathbf{l}\in \mathcal{B}^{\prime }$. In other words, one
can consistently put $A_{\mathbf{nm}}=0$ if $\mathbf{n}$ or $\mathbf{m}$ do
not belong to $\mathcal{B}^{\prime }$ without any conflict with the equation
of motion. In short, the equation of motion can be truncated to the diagonal
finite dimensional sub-matrix of $A$.

For the simplest case $n=1$, the equation of motion reduces to a scalar
relation $2\lambda _{\mathbf{n}}a_{\mathbf{nn}}+{\alpha ^{\prime }}g(a_{%
\mathbf{nn}})^{2}=0$ for $\mathbf{n}\in \mathcal{B}^{\prime }$. A
nonvanishing solution is given by $A=-\frac{2}{{\alpha ^{\prime }}g}\lambda
_{\mathbf{n}}\phi _{\mathbf{nn}}$ by using rank one projector $\phi _{%
\mathbf{nn}}$. More general diagonal solutions can be written by superposing
mutually orthogonal projectors in the subset $\mathcal{B}^{\prime }$,
\begin{eqnarray}
A_{\mathcal{B}^{\prime }} &=&-\frac{2}{{\alpha ^{\prime }}g}\sum_{\mathbf{n}%
\in \mathcal{B}^{\prime }}\lambda _{\mathbf{n}}\phi _{\mathbf{nn}}=-\frac{2}{%
{\alpha ^{\prime }}g}{\mathcal{L}_{0}}\star P_{\mathcal{B}^{\prime }}\,,
\label{proj_sol} \\
P_{\mathcal{B}^{\prime }} &=&\sum_{\mathbf{n}\in \mathcal{B}^{\prime }}\phi
_{\mathbf{nn}}\,\,,\quad P_{\mathcal{B}^{\prime }}\star P_{\mathcal{B}%
^{\prime }}=P_{\mathcal{B}^{\prime }}\,\,.
\end{eqnarray}%
We note that there are infinite number of exact and analytic solutions for
the different choices of the subset $\mathcal{B}^{\prime }$. Solution of the
form (\ref{proj_sol}) will be referred to the diagonal solutions.

Actually the solution is not restricted to the diagonal ones. To see it, one
write the equation of motion in the matrix form,
\begin{equation}
\Lambda \cdot a+a\cdot \Lambda +g{\alpha'} a\cdot a=0\,.
\end{equation}%
By shifting $a=a^{\prime }-\Lambda /({\alpha'} g)$,
the equation of motion becomes
\begin{equation}
(a^{\prime })^{2}=\Lambda ^{2}/({\alpha'}g)^{2}\,\,.
\end{equation}%
We take the subset $\mathcal{B}^{\prime }$ in such a way that for every $%
\mathbf{n}\in \mathcal{B}^{\prime }$, $\lambda _{\mathbf{n}}^{2}$ is the
same, namely the degenerate eigenspace in the right hand side of the above
equation. The equation of motion restricted to the subset $\mathcal{B}%
^{\prime }$ becomes $(a^{\prime })^{2}=\left( \lambda ^{2}
/({\alpha'}g)^{2}\right) I$
where $I$ is the identity matrix and $\lambda $ is the degenerate
eigenvalue. This equation obviously has off-diagonal solutions. As an
example, we pick $n=2$. A family of matrices which satisfy this relation is $%
A=\sum_{i=1,2,3}q_{i}\sigma _{i}$ with
$\sum_{i}q_{i}^{2}=\lambda ^{2}/({\alpha'}g)^{2}$
($\sigma _{i}$ are the Pauli matrices).

The solvability of the classical equation of motion (\ref{matrix_model})
implies there exist a similar solvability even at the quantum level if we
ignore the $\bar{x}$ dependence. We give a short comment on the analogy with
two matrix model in Appendix \ref{s:integrability}.

We would like to interpret each solution of string field theory as a new
(unstable) D-brane which is related to the original D25 brane\footnote{%
It is not obvious if the open string at zeroth order in $\gamma $ (splitting
limit) is related to D-branes. However, we use this terminology in a
generalized \textquotedblleft background\textquotedblright\ where the open
string has the frequencies $(\kappa _{e},\kappa _{o})$, which become $%
(\kappa _{e},\kappa _{e})$ when $\gamma $ is neglected.}. In order to make
this statement more explicit, we expand the action around the solution,
\begin{eqnarray}
&&S[A_{\mathcal{B}^{\prime }}+A^{\prime }]=\frac{4V}{3{\alpha ^{\prime }}%
^{3}g^{2}}\mbox{Tr}\left( {\mathcal{L}_{0}}_{\star }^{3}\star P_{\mathcal{B}%
^{\prime }}\right)   \label{S_0} \\
&&\qquad +\int d^{d}\bar{x}\,\mbox{Tr}\left( \frac{1}{2}\partial _{\bar{x}%
}A^{\prime }\star \partial _{\bar{x}}A^{\prime }+\frac{1}{{\alpha ^{\prime }}%
}{\mathcal{L}_{0}}\star (1-2P_{\mathcal{B}^{\prime }})\star A^{\prime }\star
A^{\prime }+\frac{g}{3}A^{\prime }\star A^{\prime }\star A^{\prime }\right)
\,\,.
\end{eqnarray}%
The first term (where $V$ is the volume of space-time) gives the tension of
the (un)stable D-brane
\begin{equation}
T_{\mathcal{B}^{\prime }}=\frac{4}{3{\alpha ^{\prime }}^{3}g^{2}}\sum_{%
\mathbf{n}\in \mathcal{B}^{\prime }}\lambda _{\mathbf{n}}^{3}\,\,.
\end{equation}%
The second term shows ${\mathcal{L}_{0}}$ is replaced by a new ${\mathcal{L}%
_{0}}^{\prime }$ on the (un)stable D-brane,
\begin{equation}
{\mathcal{L}_{0}}=\sum_{\mathbf{n}\in \mathcal{B}}\lambda _{\mathbf{n}}\phi
_{\mathbf{nn}}\rightarrow {\mathcal{L}_{0}}^{\prime }\equiv {\mathcal{L}_{0}}%
\star (1-2P_{\mathcal{B}^{\prime }})=\sum_{\mathbf{n}\in \mathcal{B}-%
\mathcal{B}^{\prime }}\lambda _{\mathbf{n}}\phi _{\mathbf{nn}}-\sum_{\mathbf{%
n}\in \mathcal{B}^{\prime }}\lambda _{\mathbf{n}}\phi _{\mathbf{nn}}\,.
\end{equation}%
We note that the mass squared of the matrix component $A_{\mathbf{nm}}$ is
given by the sum of the contribution from the half strings $\lambda _{%
\mathbf{n}}+\lambda _{\mathbf{m}}$. The above argument shows that the
contribution changes its sign when the label $\mathbf{n}$ is included in the
set $\mathcal{B}^{\prime }$.

\subsection{Tachyon vacuum}

Suppose we start from the theory
\begin{equation}
\lambda _{\mathbf{n}}<0\,\quad \mbox{if and only if}\quad \mathbf{n}\in
\mathcal{B}_{0}
\end{equation}%
for some subset $\mathcal{B}_{0}\subset \mathcal{B}$. In such theory, (at
least) the matrix components $A_{\mathbf{nn}}$ ($\mathbf{n}\in \mathcal{B}%
_{0}$) become the tachyonic modes. Our arguments in this section clearly
show that if we use the solution of motion $A_{\mathcal{B}_{0}}$ and
re-expand around that solution, all the negative contributions from the half
string changes sign and the tachyonic modes disappear. This is precisely the
definition of the tachyonic vacuum.

In the splitting limit, for a specific parameter choice $\kappa_e=e$ we have
$\nu=1/2$. There is only one tachyon $\mathcal{B}_0=\left\{\mathbf{0}%
\right\} $ and the tachyon vacuum becomes,
\begin{equation}  \label{tachyon-vacuum}
A=\frac{1}{{\alpha^{\prime}} g}\phi_{\mathbf{00}}\,\,.
\end{equation}
This is the butterfly state in our notation. We note that we obtain the
butterfly state as the approximate solution (namely by neglecting $S_2$) and
this is not the exact solution for the full system $S_1+S_2+S_3$.

The action around this vacuum takes the following form (after the shift of
the vacuum energy),
\begin{equation}
S[A]=\mbox{Tr}\left( \frac{1}{2}\partial_{\bar{x}}A\star \partial_{\bar{x}%
}A+ \frac{1}{{\alpha^{\prime}}} \mathcal{L}_{vac}\star A\star A+\frac{g}{3}%
A\star A\star A\right)\,, \quad \mathcal{L}_{vac}=\sum_{\mathbf{n}}|\lambda_{%
\mathbf{n}}| \phi_{\mathbf{nn}}\,,
\end{equation}
with all eigenvalues $|\lambda_\mathbf{n}|$ positive. This is the action for
the ``vacuum string field theory" in the splitting limit.

Our description of the solutions at zeroth order in $\gamma $ shares many
properties in common with the conventional VSFT proposal. One of the most
outstanding characterizations is the r\^{o}le of the projector for
describing the exact solutions of the classical equation of motion. On the
other hand, there are a few points which are different from the VSFT
proposal.

The first point is the form of the solutions. They contain the action of
Virasoro operator $-2{\mathcal{L}_{0}}\star P$ instead of the simple
projector itself as in VSFT proposal. In a sense, our solution is closer to
the solution $\Psi =Q_{L}I$ proposed in the purely cubic theory \cite{PSFT}
(after the replacement of the identity by the projector). It is due to the
fact that the kinetic term always remains in the expansion around any exact
solution.

A second point is the nature of the tachyon vacuum. As we have seen it is
characterized only by the absence of tachyonic modes in the spectrum and the
open string propagation seems to survive. Namely, the cohomology defined by
the quadratic term at the tachyonic vacuum does not appear to be trivial. In
the usual proposal, the tachyon vacuum is where there is no open string
propagation since it is the point where D-branes annihilate.

It is, of course, not very clear to which extent we should take such
\textquotedblleft discrepancies\textquotedblright\ seriously. In the Siegel
gauge there are an infinite number of subsidiary conditions that must be
applied on our solutions. We have not implemented yet these conditions. It
is likely that the ground state of the potential energy already satisfies
these conditions, but only a subset or none of the remaining extremal states
would.

A better approach to investigate this issue may be to construct the full
BRST operator in the Moyal formalism. This appears possible at $N=\infty $,
but with an infinite number of modes the issue of the midpoint is plagued
with anomalies and it is difficult to be confident that we have complete
control of the anomalies by working directly at $N=\infty .$ On the other
hand, at finite $N$ we have not figured out a substitute for the Virasoro
algebra that would be needed to construct the BRST operator. At this point
it appears quite likely that, like $L_{0},$ the full BRST operator $Q_{B}$
(a differential operator) also has a representation similar to Eq.(\ref{eom2}%
), namely%
\begin{equation}
Q_{B}A=\mathcal{Q}\star A+A\star \mathcal{Q}+qA.
\end{equation}%
We hope to report on this aspect in a future publication. Armed with such a
star product representation of $Q_{B}$ we can give a similar analysis to
what we have presented in this paper, and then we can answer the issues of
the cohomology at the tachyonic vacuum.

It is interesting to point out the following observation in relation to
closed strings. The spectrum at zeroth order in $\gamma $ (split string with
$\kappa _{o}=\kappa _{e}$) conceptually is close to the \emph{closed string}
spectrum, especially if we consider that each half string imitates the
independent modes from the left and right movers on a closed string\footnote{%
A related remark was made in \cite{Moore-Taylor}.}. This begins to give a
clue on how the graviton can be described as part of open string field
theory.

\section{Inclusion of midpoint correction}

\label{perturbation} In Witten's string field theory, the solution which
describes the tachyonic vacuum is one of the most important goal. In our
language, it corresponds to solving the equation of motion (\ref{eom})
without assuming $\gamma =0$. Since we have already solved the equation of
motion analytically in $\gamma =0$ limit, it is sensible to introduce the
effect of $\gamma $ as perturbation. For this purpose, we replace $\gamma $
by $\epsilon \gamma $ with an expansion parameter $\epsilon $. We expand $A$
as
\begin{equation}
A=A^{\left( 0\right) }+\epsilon A^{\left( 1\right) }+\epsilon ^{2}A^{\left(
2\right) }+\cdots ,  \label{expansion}
\end{equation}%
and use $A^{\left( 0\right) }$ as the solution at $\epsilon =0$. When the
spectral asymmetry between $\kappa _{e}$ and $\kappa _{o}$ is very small, we
would obtain the converging series which describe the exact solution.

We note that there is a formal analogy between our case and the analysis of
the noncommutative soliton in the scalar field theory where the equation of
motion becomes
\begin{equation}
\lbrack a,[a^{\dagger },\phi ]]+\theta V(\phi )_{\star }=0\,\,.
\end{equation}%
In the usual scenario, we first solve the second part by assuming the
parameter $\theta ,$ which is a measure of noncommutativity, is very large.
The solution is given by the noncommutative soliton as $\phi =t\phi _{0}$
where $\phi _{0}\star \phi _{0}=\phi _{0}$ and $V^{\prime }(t)=0$. The first
term is later included as the perturbation to such solutions. Then one finds
that while some solutions are stable an instability emerges for some of the
solutions \cite{soliton-stability}. Here we try to investigate our system
from a similar point of view. In our case, a similar r\^{o}le is played by
the spectral parameters $\kappa _{n}$. We can make the $\gamma $ term very
small by choosing them very close to the splitting limit.

We start from the rank 1 solution characterized by some harmonic oscillator
state labeled by $\mathbf{n}_{0}$
\begin{equation}
A^{\left( 0\right) }=-\frac{2}{\alpha' g}
\lambda _{\mathbf{n}_{0}}\phi _{\mathbf{n}_{0}\mathbf{%
n}_{0}}\,\,.
\end{equation}%
We put Eq.(\ref{expansion}) into the equation of motion and pick up $%
O(\epsilon ^{k})$ coefficients. In the first order ($k=1$) we obtain,
\begin{eqnarray}
&&{\mathcal{L}_{0}}^{\prime }\star A^{\left( 1\right) }+A^{\left( 1\right)
}\star {\mathcal{L}_{0}}^{\prime }=-\gamma A^{\left( 1\right) }\equiv
B^{\left( 1\right) }\,\,,  \label{first-order} \\
&&{\mathcal{L}_{0}}^{\prime }\equiv {\mathcal{L}_{0}}+
\alpha' g A^{\left( 0\right)
}\equiv \sum_{\mathbf{k}}\lambda _{\mathbf{k}}^{\prime }\phi _{\mathbf{kk}%
}=\sum_{\mathbf{k}}\lambda _{\mathbf{k}}(1-2\delta _{\mathbf{n}_{0}\mathbf{k}%
})\phi _{\mathbf{kk}}\,\,.
\end{eqnarray}%
The eigenvalues of the shifted ${\mathcal{L}_{0}}^{\prime }$ are exactly the
same as the modified spectrum of the half string on the unstable D-brane
which corresponds to $A^{\left( 0\right) }$ as in the previous sections. If
we expand $A^{\left( 1\right) }=\sum_{\mathbf{nm}}a_{\mathbf{nm}}^{(1)}\phi
_{\mathbf{mn}}$ and $-\gamma A^{\left( 0\right) }=\sum_{\mathbf{nm}}b_{%
\mathbf{nm}}^{(1)}\phi _{\mathbf{mn}}$, the solution to the perturbation
expansion becomes,
\begin{equation}
(\lambda _{\mathbf{n}}^{\prime }+\lambda _{\mathbf{m}}^{\prime })a_{\mathbf{%
nm}}^{(1)}=b_{\mathbf{nm}}^{(1)}\,\,.  \label{Loprime}
\end{equation}%
This equation has a unique solution as long as $(\lambda _{\mathbf{n}%
}^{\prime }+\lambda _{\mathbf{m}}^{\prime })\neq 0$. We note that $(\lambda
_{\mathbf{n}}^{\prime }+\lambda _{\mathbf{m}}^{\prime })$ gives the mass
squared of the open string on the (unstable) D-brane. The recursion relation
breaks if there exist massless excitations. Such modes exist if (i)there
exists $\mathbf{m}(\neq \mathbf{n}_{0})$ such that $\lambda _{\mathbf{m}%
}=\lambda _{\mathbf{n}_{0}}$ (we suppose $\lambda _{\mathbf{n}_{0}}\neq 0$).
or (ii) $\lambda _{\mathbf{n}}^{\prime }=0$ for some $\mathbf{n}$. In such
situations, we need to impose $b_{\mathbf{nm}}^{(1)}=b_{\mathbf{mn}}^{(1)}=0$
in order to have a perturbation expansion with a nontrivial solution.
However, this type of constraint becomes rather nontrivial when we need to
solve the higher order equation.

For the situation (i), we believe that the rank 1 solutions become singular
in the perturbation series. One resolution for the degenerate case is to
consider the higher rank projector and only begin with the solution of the
form $A^{\left( 0\right) }=-\frac{2\lambda}{\alpha' g}
\sum_{i\in \Lambda }\phi _{ii}$ where $%
\Lambda $ is the set of indices with $\lambda _{i}=\lambda $. Starting from
this solution, it is not possible to have $\lambda _{\mathbf{n}}^{\prime
}+\lambda _{\mathbf{m}}^{\prime }=0,$ and the recursion formula becomes
consistent and have a unique solution. For the situation (ii), there does
not seem to exist such a cure. One possibility is, however, to shift the
splitting of $L_{0}$ into ${\mathcal{L}_{0}}$ and $\gamma $ slightly, ${%
\mathcal{L}_{0}}\rightarrow {\mathcal{L}_{0}}+b$, $\gamma \rightarrow \gamma
-2b$. This shifts the eigenvalues of ${\mathcal{L}_{0}}$ by a constant and
escape the singularity mentioned above\footnote{%
We have to mention that such a dangerous situation seems to appear at least
naively. In the open string limit (\ref{open_string_limit}), the vacuum
energy $\nu $ defined in (\ref{nu_and_cL}) becomes divergent. If we use the
zeta function regularization to obtain a finite value for $\nu $, we need to
use,
\begin{equation}
\sum_{e>0}e-\sum_{o>0}o=2(\zeta (-1)-\zeta (-1,1/2))=2\left( -\frac{1}{24}-%
\frac{1}{12}\right) =-\frac{1}{4}\,\,.  \label{zeta-reg}
\end{equation}%
For the critical dimension $d=26$, it makes $\nu =2$. With such a choice for
$\nu $, there exists a \textquotedblleft graviton-like\textquotedblright\
excitation $\beta _{2}^{\mu \dagger }\star \phi _{\mathbf{00}}\star \beta
_{2}^{\nu }$ which becomes exactly massless. Furthermore one can easily
check the right-hand side of (\ref{first-order}) is also nonvanishing
(dilaton-like excitation),
\begin{equation}
B_{1}\propto \frac{1}{1+\bar{w}w}(w_{2})^{2}\eta _{\nu \mu }\beta _{2}^{\nu
\dagger }\star \phi _{\mathbf{00}}\star \beta _{2}^{\mu }+\cdots .
\end{equation}%
The situation is, however, very delicate. If we take the naive open string
limit $\bar{w}w\rightarrow \infty $, this term also vanishes. This is the
usual problem of taking the naive limit. The proposal in MSFT \cite{BM1}\cite%
{BM2}\cite{BKM1} is to use the finite $N$ regularization in all the
intermediate computation and take the large $N$ limit only at the end of the
calculation. The divergence which we encounter is caused by the use of the
zeta-function regularization (\ref{zeta-reg}) at the intermediate step of
the computation which becomes quite dangerous. The correct prescription will
be to take $\nu $ unfixed and solve the recursion and only take the limit (%
\ref{zeta-reg}) after we sum over all the perturbation expansion.}.

As an explicit example, we present the first order correction if we take the
butterfly state as the zeroth term $A_{0}=\frac{2}{{\alpha ^{\prime }}g}\nu
\phi _{\mathbf{00}}$. We use the oscillator representation of the action of
gamma terms to the string fields,
\begin{eqnarray}
\gamma ^{matter}A &=&\frac{-1}{8(1+\bar{w}w)}\sum_{e,e^{\prime }>0}\sqrt{%
\kappa _{e}}w_{e}\sqrt{\kappa }_{e^{\prime }}w_{e^{\prime }}\left\{ \beta
_{e}^{\mu }+\beta _{e}^{\mu \dagger },\left\{ \beta _{e^{\prime }}^{\nu
}+\beta _{e^{\prime }}^{\nu \dagger },A\right\} _{\star }\right\} _{\star
}\eta _{\mu \nu }\,, \\
\gamma ^{gh}A &=&\frac{1}{4(1+\bar{w}w)}\sum_{e,e^{\prime }>0}w_{e}\kappa
_{e}w_{e^{\prime }}\left\{ \beta _{e}^{c}-\beta _{e}^{c\dagger },\left[
\beta _{e^{\prime }}^{b}+\beta _{e^{\prime }}^{b\dagger },A\right] _{\star
}\right\} _{\star }\,\,.
\end{eqnarray}%
The first order correction is given as,
\begin{eqnarray}
{\alpha ^{\prime }}gA_{1} &=&\frac{(d-2)\delta }{4}\,\phi _{\mathbf{00}}+%
\frac{\nu }{4(1+\bar{w}w)}\sum_{e,e^{\prime }>0}\left( \frac{%
w_{e}w_{e^{\prime }}}{\kappa _{e}+\kappa _{e^{\prime }}}D_{ee^{\prime }}^{1}+%
\frac{2w_{e}w_{e^{\prime }}}{\kappa _{e}+\kappa _{e^{\prime }}-2\nu }%
D_{ee^{\prime }}^{2}\right) \,\,, \\
\delta &\equiv &\frac{\sum_{e>0}\kappa _{e}w_{e}^{2}}{1+\bar{w}w}\,\,, \\
D_{ee^{\prime }}^{1} &=&\eta _{\mu \nu }\sqrt{\kappa _{e}\kappa _{e^{\prime
}}}\left( \beta _{e}^{\mu \dagger }\star \beta _{e^{\prime }}^{\nu \dagger
}\star \phi _{\mathbf{00}}+\phi _{\mathbf{00}}\star \beta _{e^{\prime
}}^{\nu }\star \beta _{e}^{\mu }\right)  \notag \\
&&+2\kappa _{e}\left( \beta _{e}^{c\dagger }\star \beta _{e^{\prime
}}^{b\dagger }\star \phi _{\mathbf{00}}+\phi _{\mathbf{00}}\star \beta
_{e^{\prime }}^{b}\star \beta _{e}^{c}\right) \,\,, \\
D_{ee^{\prime }}^{2} &=&\eta _{\mu \nu }\sqrt{\kappa _{e}\kappa _{e^{\prime
}}}\beta _{e}^{\mu \dagger }\star \phi _{\mathbf{00}}\star \beta _{e^{\prime
}}^{\nu }-\kappa _{e}\left( \beta _{e}^{c\dagger }\star \phi _{\mathbf{00}%
}\star \beta _{e^{\prime }}^{b}+\beta _{e^{\prime }}^{b\dagger }\star \phi _{%
\mathbf{00}}\star \beta _{e}^{c}\right) \,\,.
\end{eqnarray}%
We note that the first order correction is small compared with the zeroth
order term if $\bar{w}w<\!<1$, namely in the vicinity of the splitting limit
(\ref{splitting}). On the other hand, in the open string limit (\ref%
{open_string_limit}), while the combination $w_{e}/\sqrt{1+\bar{w}w}$
becomes very small (which are the coefficients of $D_{1,2}$), $\bar{w}w$, $%
\delta $ and $\nu $ are naively divergent. In this sense, the applicability
of the perturbation series in the open string limit seems to be quite subtle.

One possibility to overcome this difficulty is to use the ambiguity of the
splitting of ${\mathcal{L}_{0}}$ and $\gamma $ which is mentioned in section
\ref{sub:splitting}.
With some careful choice, for example, it seems that one can remove
the divergence in $\nu $ and $\delta $ which appear at the first order. The
problem of higher corrections, however, is very delicate and we would like
to postpone the careful treatment of these problems to a future publication.

Due to the correction to the tachyon vacuum, the formula for the brane
tension should also be modified. We expand the action in the following form,
\begin{equation}
S[A^{\left( 0\right) }+\epsilon A^{\left( 1\right) }+\epsilon ^{2}A^{\left(
2\right) }+\cdots ]=S^{\left( 0\right) }+\epsilon S^{\left( 1\right)
}+\epsilon ^{2}S^{\left( 2\right) }+\cdots .
\end{equation}%
If we start the perturbation series around the solution $A^{\left( 0\right)
}=-\frac{2}{{\alpha ^{\prime }}g}{\mathcal{L}_{0}}\star P$ (with ${\mathcal{L%
}_{0}}\star P=P\star {\mathcal{L}_{0}}$, $P\star P=P$), the zeroth term is
given in Eq.(\ref{S_0}). The first order correction is given by
\begin{equation}
S^{\left( 1\right) }=-\frac{2}{{\alpha ^{\prime }}^2g}
V\mbox{Tr}\,\left( {%
\mathcal{L}_{0}}_{\star }^{2}\star P\star A^{\left( 1\right) }\right) \,\,.
\end{equation}%
In the perturbation around the butterfly state, we evaluate the tension as,%
\footnote{%
This will be compared to the ordinary D25-brane tension after a self
consistent normalization of the action in MSFT\cite{PREP}.}
\begin{equation}
T=\frac{1}{{\alpha ^{\prime }}^{3}g^{2}}\left( -\frac{4}{3}\nu ^{3}-\epsilon
\frac{\nu ^{2}\delta }{2}(d-2)\right) +O(\epsilon ^{2})\,\,.
\end{equation}

We can continue the perturbation expansion for higher $k$. The recursion
formula is already given in (\ref{recursion}).
With the above redefinition of $A_{0}$ for the degenerate case, we can solve
this equation term by term for any spectrum uniquely. The second order
perturbation is, for example, given as, 
\begin{eqnarray}
A^{\left( 2\right) } &=&(-L_{0}^{\prime })^{-1}\left( \gamma A^{\left(
1\right) }\right) +{\alpha ^{\prime }}g(-L_{0}^{\prime })^{-1}(A^{\left(
1\right) }\star A^{\left( 1\right) })\,\,, \\
&&\quad \mbox{with}\quad A^{\left( 1\right) }=(-L_{0}^{\prime })^{-1}\left(
\gamma A^{\left( 0\right) }\right) \,\,.
\end{eqnarray}%
where $L_{0}^{\prime }$ applied on any field is defined by $L_{0}^{\prime
}A\equiv {\mathcal{L}_{0}}^{\prime }\star A+A\star {\mathcal{L}_{0}}^{\prime
},$ while $\left( L_{0}^{\prime }\right) ^{-1}$ on any field is given by $%
\left( L_{0}^{\prime }\right) ^{-1}A=\int_{0}^{\infty }d\tau ~e_{\star
}^{-\tau {\mathcal{L}_{0}}^{\prime }}\star A\star e_{\star }^{-\tau {%
\mathcal{L}_{0}}^{\prime }},$ as explained in Eq.(\ref{int}) and footnote (%
\ref{intEq}). More directly, both $L_{0}^{\prime }$ and $\left(
L_{0}^{\prime }\right) ^{-1}$ are simple algebraic expressions in the matrix
notation of Eqs.(\ref{first-order}-\ref{Loprime}).

Actually one may obtain a formal expression for $A$ for the entire
perturbation sum. For that purpose, we introduce the \textquotedblleft
dressed propagator\textquotedblright ,
\begin{equation}
(-L_{0}^{\prime })^{-1}+(-L_{0}^{\prime })^{-1}\epsilon \gamma
(-L_{0}^{\prime })^{-1}+(-L_{0}^{\prime })^{-1}\epsilon \gamma
(-L_{0}^{\prime })^{-1}\epsilon \gamma (-L_{0}^{\prime })^{-1}\cdots
=(-L_{0}^{\prime }-\epsilon \gamma )^{-1},
\end{equation}%
and \textquotedblleft dressed version\textquotedblright\ of $A^{\left(
1\right) }$ and star product,
\begin{equation}
\tilde{A}^{\left( 1\right) }\equiv (-L_{0}^{\prime }-\epsilon \gamma
)^{-1}\gamma A^{\left( 0\right) }\,,\quad A\bullet B\equiv {\alpha ^{\prime }%
}g(-L_{0}^{\prime }-\epsilon \gamma )^{-1}(A\star B)\,\,.
\end{equation}%
We note that the \textquotedblleft bullet product $\bullet $%
\textquotedblright\ is not associative product. One may then claim that full
wave function $A$ can be expressed as,
\begin{eqnarray}
A &=&A^{\left( 0\right) }+\epsilon \tilde{A}^{\left( 1\right) }+\epsilon ^{2}%
\tilde{A}^{\left( 1\right) }\bullet \tilde{A}^{\left( 1\right) }+\epsilon
^{3}\left( (\tilde{A}^{\left( 1\right) }\bullet \tilde{A}^{\left( 1\right)
})\bullet \tilde{A}^{\left( 1\right) }+\tilde{A}^{\left( 1\right) }\bullet (%
\tilde{A}^{\left( 1\right) }\bullet \tilde{A}^{\left( 1\right) })\right)
+\cdots  \notag \\
&=&A^{\left( 0\right) }+\sum_{n=1}^{\infty }\epsilon ^{n}\left(
\mbox{(all possible associations
of) }\underbrace{\tilde{A}^{\left( 1\right) }\bullet \cdots \bullet \tilde{A}%
^{\left( 1\right) }}_{n}\right) \,\,.  \label{full_sols}
\end{eqnarray}%
For example, $\epsilon ^{4}$ term is given as,
\begin{eqnarray}
&&\tilde{A}^{\left( 1\right) }\bullet (\tilde{A}^{\left( 1\right) }\bullet (%
\tilde{A}^{\left( 1\right) }\bullet \tilde{A}^{\left( 1\right) }))+(\tilde{A}%
^{\left( 1\right) }\bullet \tilde{A}^{\left( 1\right) })\bullet (\tilde{A}%
^{\left( 1\right) }\bullet \tilde{A}^{\left( 1\right) })+\tilde{A}^{\left(
1\right) }\bullet ((\tilde{A}^{\left( 1\right) }\bullet \tilde{A}^{\left(
1\right) })\bullet \tilde{A}^{\left( 1\right) })  \notag \\
&&+(\tilde{A}^{\left( 1\right) }\bullet (\tilde{A}^{\left( 1\right) }\bullet
\tilde{A}^{\left( 1\right) }))\bullet \tilde{A}^{\left( 1\right) }+((\tilde{A%
}^{\left( 1\right) }\bullet \tilde{A}^{\left( 1\right) })\bullet \tilde{A}%
^{\left( 1\right) })\bullet \tilde{A}^{\left( 1\right) }.  \notag
\end{eqnarray}%
A proof of the formula (\ref{full_sols}) is given by the use of the
recursion formula (\ref{recursion}). An easier proof is to write down the
equation of motion for the deviation $A^{\prime }\equiv A-A^{\left( 0\right)
}=\epsilon A^{\left( 1\right) }+\epsilon ^{2}A^{\left( 2\right) }+\cdots $,
\begin{equation}
-(L_{0}^{\prime }+\epsilon \gamma )A^{\prime }=\epsilon \gamma A^{\left(
0\right) }+{\alpha ^{\prime }}gA^{\prime }\star A^{\prime }\,\,.
\end{equation}%
{}From above definitions, one may rewrite it as, $A^{\prime }=\epsilon
\tilde{A}^{\left( 1\right) }+A^{\prime }\bullet A^{\prime }$. We use this
relation recursively to obtain (\ref{full_sols}), for example,
\begin{equation}
A^{\prime }=\epsilon \tilde{A}^{\left( 1\right) }+(\epsilon \tilde{A}%
^{\left( 1\right) }+A^{\prime }\bullet A^{\prime })\bullet (\epsilon \tilde{A%
}^{\left( 1\right) }+A^{\prime }\bullet A^{\prime })=\cdots \,\,.
\end{equation}

With this explicit formula, we claim that there exists a unique solution (%
\ref{full_sols}) to the full string field equation for each solution in the
splitting limit as long as the perturbation expansion is convergent. We hope
that the careful analysis in the vicinity of the splitting limit will reveal
some nature of the dynamics in the open string limit.

We note that the solutions becomes the projector with respect to Witten's
star product $\star$ only at the splitting limit and the perturbation breaks
such simplicity.

\section{Conclusion}

We have seen in this paper that the splitting limit gives a system
where the translational invariant solutions are solved
analytically in terms of the projection operators. We argued that
this is an analog of the large $\theta $ limit in the
noncommutative scalar field theory. We can introduce the midpoint
effect as a perturbation series which is analogous to the finite
$\theta $ case.
In the development of open string field theory, this gives the
first example where the r\^{o}le of noncommutative solitons is
explicitly demonstrated with a careful treatment of the midpoint
correction. We believe that it gives a firm ground upon which the
relation between noncommutative geometry and open string field
theory will be discussed in the future.

There are, of course, many topics which should be clarified in a
future study. One of the most interesting directions is to find
the analytic solution for $\gamma \neq 0$ in a closed form. While
this appears difficult because we cannot find a basis which
diagonalizes $S_{i}$ ($i=1,2,3$) simultaneously, there may be a
possibility that a few of the exact solutions, in particular the
true vacuum, could be derived with some insight.

In our description, the tachyon vacuum still seems to have an open
string spectrum (modulo the extra gauge invariance conditions in
the Siegel gauge). Elimination of these modes would be possible
only when the exact solution is found in a closed form, and the
remaining gauge invariance conditions are imposed.

A related issue is the BRST symmetry. So far with arbitrary choice
of the spectrum $\kappa _{n}$ as a function of $n$, we cannot
define the nilpotent BRST operator. The merit of our approach is
that one can handle the midpoint correction at finite $N$. While
the BRST charge exists in the open string limit at $\kappa
_{n}=n$, it is very challenging to see how the BRST operator will
be affected in that limit by the midpoint correction we have
emphasized.

Another essential question is whether the splitting limit itself
can be interpreted as a ``real string'' defined by some kind of
(B)CFT. As an example, we take $\kappa_e=e$ and $N=\infty$. The
spectrum of the model is generated by two set of oscillators
$a^\dagger_e$ (and $a_e$) action on the vacuum state from left
(and right). In a sense each of two half string behaves exactly
like the original open string. This is somewhat similar to the
closed string excitation where two sets of oscillators are left
and right moving modes. Since there is no $L_0-\bar{L}_0=0$
constraint in the splitting limit, it is certainly different from
the closed string. However this analogy may have some implication
of the nature of the tachyon vacuum where we are supposed to have
only the closed string excitations.

While it is more speculative, we may comment on the relation with
D-branes. Usually in BCFT, D-branes are described by boundary
states in the closed string Hilbert space. We note that the open
string Hilbert space in the splitting limit has a similar
structure as the closed string. The similarity may imply that the
deformation of the open string Hilbert space is needed to describe
the D-brane as a projector in the open string Hilbert space. This
kind of comment might be helpful if we want to perform a similar
analysis in a generic closed string background.

\begin{center}
\textbf{Acknowledgment}
\end{center}

I.B. is supported in part by a DOE grant DE-FG03-84ER40168. I.K. is
supported in part by JSPS Research Fellowships for Young Scientists. Y.M. is
supported in part by Grant-in-Aid (\# 13640267) from the Ministry of
Education, Science, Sports and Culture of Japan.

\appendix

\section{Definitions in MSFT}

We review some basic definitions in the Moyal star formulation of
string field theory \cite{B,BM1,BM2,BKM1}.


To provide a regulator in MSFT, we use an explicit truncation of the number
of oscillators ($1\leq |n|\leq 2N$) and introduce the $2N$ arbitrary
\textquotedblleft frequency\textquotedblright\ parameters $\kappa _{n}$ ($%
n=1,\cdots ,2N$). These appear in the commutation relations among the
oscillators, such as
\begin{equation}
\lbrack \alpha _{n}^{\mu },\alpha _{n^{\prime }}^{\nu }]=\kappa _{n}^{\prime
}\delta _{n+n^{\prime }}\eta ^{\mu \nu }\,\,,\quad \kappa _{n}^{\prime
}\equiv \epsilon (n)\kappa _{|n|}\,\,.  \label{osc_comm}
\end{equation}%
In the following we need to distinguish the frequency for even and odd
labels, and write them as $\kappa _{e}^{\prime }$, $\kappa _{o}^{\prime },$
where $e$ (resp. $o$) runs over even (resp. odd) numbers in the range of $n$%
.

In the definition of the canonical variables in Moyal space, there is a
Bogoliubov transformation $U_{-e,o}$ from the oscillators labeled with odd
numbers $o$ to those labeled with even numbers $e$. In \cite{BM1, BM2, BKM1}%
, $U$ is related to a set of special matrices and vectors $T,R,S,v^{\prime
},w^{\prime }$. These are all functions of the frequency parameters $\kappa
_{e},\kappa _{o}.$ In the limit (which we will refer to as \textquotedblleft
the open string limit\textquotedblright )
\begin{equation}
N\rightarrow \infty \,,\quad \kappa _{o}\rightarrow o\,,\quad \kappa
_{e}\rightarrow e\,,  \label{open_string_limit}
\end{equation}%
the basic matrix $U$ and vectors $w_{e}^{\prime },v_{e}^{\prime }$ (for both
positive and negative integers $e,o$) are
\begin{equation}
U_{-e,o}\rightarrow \frac{2}{\pi }\frac{i^{o-e-1}}{o-e}\,,\quad
w_{e}^{\prime }\rightarrow i^{-e+2}\,,\quad v_{o}^{\prime }\rightarrow \frac{%
2}{\pi }\frac{\,i^{o-1}}{o}\quad \mbox{and }\ \bar{w}^{\prime }w^{\prime
}\rightarrow \infty .  \label{limit}
\end{equation}%
Compared to the notation $w_{e},v_{o}$ that we also use for positive
integers, $w^{\prime }$ and $v^{\prime }$ are defined as
\begin{equation}
w_{e}^{\prime }=w_{|e|}/\sqrt{2},\;\;v_{o}^{\prime }=v_{|o|}/\sqrt{2}.
\end{equation}

In the regulated version of MSFT with finite $N$ these matrices are deformed
as functions of arbitrary $\kappa _{e},\kappa _{o}$ as follows
\begin{eqnarray}
&&U_{-e,o}=\frac{w_{e}^{\prime }v_{o}^{\prime }\kappa _{o}^{\prime }}{\kappa
_{e}^{\prime }-\kappa _{o}^{\prime }}\,,\quad U_{-o,e}^{-1}=\frac{%
w_{e}^{\prime }v_{o}^{\prime }\kappa _{e}^{\prime }}{\kappa _{e}^{\prime
}-\kappa _{o}^{\prime }}\,,\quad UU^{-1}=U^{-1}U=1\,\,,  \label{U_exp} \\
&&w_{e} ={i^{2-e}}\frac{\prod_{o^{\prime }>0}\left\vert \kappa
_{e}^{2}/\kappa _{o^{\prime }}^{2}-1\right\vert ^{\frac{1}{2}}}{%
\prod_{e^{\prime }(\neq e)>0}\left\vert \kappa _{e}^{2}/\kappa _{e^{\prime
}}^{2}-1\right\vert ^{\frac{1}{2}}},\quad v_{o}={i^{o-1}}\frac{%
\prod_{e^{\prime }>0}\left\vert 1-\kappa _{o}^{2}/\kappa _{e^{\prime
}}^{2}\right\vert ^{\frac{1}{2}}}{\prod_{o^{\prime }(\neq o)>0}\left\vert
1-\kappa _{o}^{2}/\kappa _{o^{\prime }}^{2}\right\vert ^{\frac{1}{2}}}
\,\,,
\label{wv_exp} \\
&&T_{eo}=U_{-e,o}+U_{e,o},\quad
R_{oe}=U_{-o,e}^{-1}+U_{o,e}^{-1},\quad
S_{eo}=U_{-e,o}-U_{e,o}=U_{-o,e}^{-1}-U_{o,e}^{-1}\,,
\end{eqnarray}%
where a bar on a matrix means its transpose. Of course, these expressions
reduce to their limiting values in Eq.(\ref{limit}) in the large $N$ limit.
They satisfy the following relations for arbitrary $\kappa _{n}$ (including
the limit of Eq.(\ref{open_string_limit}))
\begin{eqnarray}
&& U^{-1}={\kappa _{o}^{\prime }}^{-1}\bar{U}\kappa _{e}^{\prime }=\bar{U}%
+v^{\prime }\bar{w}^{\prime }\,,\quad U\bar{U}=1-\frac{w^{\prime }\bar{w}%
^{\prime }}{1+\bar{w}^{\prime }w}\,,\quad v^{\prime
}=\bar{U}w^{\prime }\,\,\label{relations} \\
&& TR=RT=1\,,\quad S\bar{S}=\bar{S}S=1\,\,,\;\;T=\kappa
_{e}^{-1}S\kappa _{o},\;R=\kappa _{o}^{-1}\bar{S}\kappa _{e}.
\end{eqnarray}%
In Eq.(\ref{relations}) the first formula implies that $U$ changes
the spectrum from $\kappa _{o}^{\prime }$ to $\kappa _{e}^{\prime
},$ whereas the second one gives the origin of the midpoint
correction. The $v_{o}^{\prime }$ and $w_{e}^{\prime } $ vectors
are related to each other through the third relation. This is only
a partial list of relations; for the complete set of relations see \cite{BM2}%
.

In most explicit computations it is much more efficient to use the relations
among the matrices rather than their explicit messy expressions. We
emphasize that these relations hold for any values of $\kappa _{n},$
including those of the limiting case in Eq.(\ref{open_string_limit}).
Therefore, performing analytic computations at finite $N$ is often not any
more difficult than performing them at infinite $N.$

In the Moyal language, the action of the oscillators $\alpha $ on the string
field is represented by the action of $\beta $ oscillators defined in Eq.(%
\ref{beta-osc}). Suppose the oscillator state $|\psi \rangle $ corresponds
to Moyal field $A$ (in Siegel gauge),
\begin{eqnarray}
\alpha _{e}^{\mu }|\psi \rangle  &\longleftrightarrow &\hat{\beta}_{e}^{\mu
}A\equiv \sqrt{\frac{\kappa _{e}}{2}}(\beta _{e}^{\mu }\star A-A\star \beta
_{-e}^{\mu })-w_{e}^{\prime }\beta _{0}^{\mu }A\,\,, \\
\alpha _{o}^{\mu }|\psi \rangle  &\longleftrightarrow &\hat{\beta}_{o}^{\mu
}A\equiv \sum_{e\neq 0}(\bar{\beta}_{e}^{\mu }A)U_{-e,o}=\sqrt{\frac{\kappa
_{e}}{2}}(\beta _{o}^{\mu }\star A+A\star \beta _{-o}^{\mu })\,, \\
b_{e}|\psi \rangle  &\longleftrightarrow &\hat{\beta}_{e}^{b}A\equiv \frac{1%
}{\sqrt{2}}\left( \beta _{e}^{b}\star A+(-1)^{|A|}\star \beta
_{-e}^{b}\right)\,\,,  \\
b_{o}|\psi \rangle  &\longleftrightarrow &\hat{\beta}_{o}^{b}A\equiv
\sum_{e\neq 0}(\bar{\beta}_{e}^{b}A)U_{-e,o}\equiv \frac{1}{\sqrt{2}}\left(
\beta _{o}^{b}\star A-(-1)^{|A|}A\star \beta _{-o}^{b}\right)
\,\,,\\
c_{e}|\psi \rangle  &\longleftrightarrow &\hat{\beta}_{e}^{c}A\equiv \frac{1%
}{\sqrt{2}}\left( \beta _{e}^{c}\star A-(-1)^{|A|}A\star \beta
_{-e}^{c}\right) \,\,, \\
c_{o}|\psi \rangle  &\longleftrightarrow &\hat{\beta}_{o}^{c}A\equiv
\sum_{e\neq 0}(U_{o,-e}^{-1}\bar{\beta}_{e}^{c})A\equiv \frac{1}{\sqrt{2}}%
\left( \beta _{o}^{c}\star A+(-1)^{|A|}\star \beta _{-o}^{c}\right)
\end{eqnarray}%
where $\beta _{o}$ is Bogoliubov transformation of $\beta _{e}$, namely
\begin{equation}
\beta _{o}^{\mu }=\sum_{e\neq 0}\beta _{e}^{\mu }U_{-e,o}\,\,,
\quad\beta
_{o}^{b}=\sum_{e\neq 0}\beta _{e}^{b}U_{-e,o}\,\,,\quad\beta
_{o}^{c}=\sum_{e\neq 0}U_{o,-e}^{-1}\beta _{e}^{c}\,\,.
\end{equation}%
Note that $\hat{\beta}_{e,o}^{\mu ,b,c},\bar{\beta}_{e,o}^{\mu ,b,c}$ are
differential operators, but $\beta _{e,o}$ are fields multiplied with the
Moyal star. We can prove that the even differential operators $\hat{\beta}%
_{e},\,\bar{\beta}_{e}$ satisfy the standard
(anti-) commutation relation for the
even mode oscillators,
\begin{equation}
\left[ \hat{\beta}_{e}^{\mu },\hat{\beta}_{e^{\prime }}^{\nu }\right] =\left[
\bar{\beta}_{e}^{\mu },\bar{\beta}_{e^{\prime }}^{\nu }\right] =\eta ^{\mu
\nu }\kappa _{e}^{\prime }\delta _{e+e^{\prime }}\,,\quad \left\{ \hat{\beta}%
_{e}^{b},\hat{\beta}_{e^{\prime }}^{c}\right\} =\left\{ \bar{\beta}_{e}^{b},%
\bar{\beta}_{e^{\prime }}^{c}\right\} =\delta _{e+e^{\prime }}\,\,.
\end{equation}%
On the other hand, after Bogoliubov transformation, $\hat{\beta}_{o}^{p}$
satisfies the odd mode commutation relation,
\begin{equation}
\left[ \hat{\beta}_{o}^{\mu },\hat{\beta}_{o^{\prime }}^{\nu }\right]
=\kappa _{o}^{\prime }\delta _{o+o^{\prime }}\,,\quad \left\{ \hat{\beta}%
_{o}^{b},\hat{\beta}_{o^{\prime }}^{c}\right\} =\delta _{o+o^{\prime }}\,\,.
\end{equation}

{}For arbitrary frequencies, one can define the perturbative string states.
In particular, the perturbative vacuum in the oscillator language is mapped
to the gaussian function \cite{BM2,PREP},
\begin{eqnarray}
&&A_{0}\sim \exp \left( -\bar{\xi}^{\mu }M_{0}\xi _{\mu }-2\bar{\xi}%
^{b}M_{0}^{gh}\xi ^{c}\right) ,\quad \bar{\xi}^{b}=
(\bar{x}_{e}^{b}\,\,\bar{p}%
_{e}^{b}),\,\quad \bar{\xi}_{2}^{c}=(\bar{x}_{e}^{c}\,\bar{p}_{e}^{c})\,, \\
&&M_{0}=\left(
\begin{array}{cc}
\frac{\kappa _{e}}{2l_{s}^{2}} & 0 \\
0 & \frac{2l_{s}^{2}}{\theta ^{2}}T\kappa _{o}^{-1}\bar{T}%
\end{array}%
\right) ,\quad M_{0}^{gh}=\left(
\begin{array}{cc}
\frac{i}{2}\bar{R}\kappa _{o}R & 0 \\
0 & \frac{2i}{{\theta ^{\prime }}^{2}}\kappa _{e}^{-1}%
\end{array}%
\right) \,\,.  \label{vac}
\end{eqnarray}%
Similarly one can construct the Moyal map of coherent states which
correspond to adding a linear term in the exponent of the gaussian above.
With this setup, the Moyal star is used to compute generalizations of
Neumann coefficients. It is shown \cite{BM2} that they satisfy the basic
nonlinear relations given by Gross and Jevicki even for arbitrary
frequencies. The generalized Neumann matrices $\left( V_{n}^{\left[ rs\right]
}\right) _{kl},\left( V_{n}^{\left[ rs\right] }\right) _{k0},\left( V_{n}^{%
\left[ rs\right] }\right) _{00}$ for any $n$-string vertex are
shown to be simple explicit functions of the single matrix
$t_{eo}=\kappa _{e}^{1/2}T_{eo}\kappa _{o}^{-1/2}.$ Diagonalizing
this single matrix diagonalizes simultaneously all Neumann
matrices. This explains and justifies the notion of Neumann
spectroscopy for arbitrary oscillator frequencies. These results
were initially obtained in the matter sector (or with bosonized
ghosts) but by now they have been generalized to include also the
fermionic ghost sector \cite{PREP}. In the open string limit we
fix the frequencies to Eq.(\ref{open_string_limit}). In this limit
our generalized Neumann matrices agree with other computations of
these coefficients.

Furthermore, for arbitrary frequencies $\kappa _{n}$, using the Moyal star,
one can also compute open string amplitudes \cite{BKM1} including the ghost
sector \cite{PREP}.

The regulator is removed by taking the limit in
Eq.(\ref{open_string_limit}) at the end of computations. As
emphasized in \cite{BKM1}, taking such a limit at the Lagrangian
level is wrong because of anomalies and leads to inconsistent
results. We note that we break the conformal symmetry explicitly
when we work at finite $N$ or arbitrary values of $\kappa _{n}$.
This is the cost to pay to resolve the associativity anomaly among
the basic relations. We expect that the
conformal symmetry is re-established in the limit of Eq.(\ref%
{open_string_limit}).

It has been shown that this regularization scheme gives the
correct results in explicit computations, including the spectrum
of $L_{0},$ perturbative states, Neumann coefficients, string
Feynman graphs, and numerical estimates of certain quantities
computed with other methods in the literature.


\section{Solvability of $S_2+S_3$}

\label{s:particle_limit}
In this appendix, we show that the combination $S_2+S_3$ is also solvable as
in the combination $S_1+S_3$ considered in the text. For simplicity we
consider only the matter sector and keep just one space-time component. We
change variable (symplectic transformation) from $x_e,p_e$ to $y_e,q_e$ such
that $y_2=\frac{1}{\sqrt{1+\bar w w}}\sum_e w_e p_e$. The e.o.m. becomes $%
q_2^2A+{\alpha^{\prime}} g A\star A=0$ but since $q_4, q_6,\cdots$ and $y_4,
y_6,\cdots$ are irrelevant, we wrote it simply as, (neglecting subscript 2),
\begin{equation}
q^2A(y,q)+{\alpha^{\prime}} g (A\star A)(y,q)=0\,\,.
\end{equation}
While the first term does not split as before, one may always write it as,
\begin{equation}
\frac{1}{4}\left\{q,\left\{q,A\right\}_\star\right\}_\star +{\alpha^{\prime}}
g A\star A=0\,\,.
\end{equation}
While the kinetic term does not split, we may use the same trick as before
to write down one family of solutions. For that purpose, we prepare the wave
function which is diagonal with respect to $q$,
\begin{eqnarray}
&&q\star\phi(k,l)=k \,\phi(k,l),\qquad \phi(k,l)\star q=l
\,\phi(k,l)\,\qquad k,l\in{\mathbf{R}} \,, \\
&& \phi(k,l)\star
\phi(k^{\prime},l^{\prime})=\delta(l-k^{\prime})\,\phi(k,l^{\prime})\,\,.
\end{eqnarray}
The solution to this definition is given as,
\begin{equation}
\phi(k,l)=\delta\left(q-(k+l)/{2}\right)\,e^{i(k-l)y}\,\,.
\end{equation}
If we expand $A=\int dk dl A(k,l) \phi(l,k)$, e.o.m. can be rewritten in
terms of $A(k,l)$ as
\begin{equation}
\left(\frac{k+l}{2}\right)^2 A(k,l)+{\alpha^{\prime}} g \int dr
A(k,r)A(r,l)=0\,\,.
\end{equation}
A family of solutions which is similar to those given in the previous
section can be written as,
\begin{equation}
A_\theta(k,l) = -\frac{k^2}{{\alpha^{\prime}} g } \theta_\Sigma (k)
\delta(k-l)
\end{equation}
with
\begin{equation}
\theta_\Sigma(k) =\left\{%
\begin{array}{ll}
1 \qquad & k\in \Sigma \\
0 \qquad & \mbox{otherwise}%
\end{array}%
\right.
\end{equation}
where $\Sigma$ is a certain range in \textbf{R}. We note that the
projector $\phi(k,k)=\delta(q-k)$ does not depend on the coordinate $y$.

The tension is computed similarly as
\begin{equation}
S[A_\Sigma]=\frac{V}{6{\alpha^{\prime}}^3g^2} \int_\Sigma dk \,k^6\,\,.
\end{equation}
The volume factor $V$ appears here because $A$ does not depend on the
coordinate. Since the projector is defined over the continuum variable, we
expect any nontrivial solution obtained here will be unstable except for the
trivial one $A=0$\,\,.


\section{Integrability of the matrix model}

\label{s:integrability} The fact that we can solve the translational
invariant solutions of Eq.(\ref{matrix_model}) implies that it is also
integrable even at the quantum level as long as we neglect the $\bar{x}$
dependence (namely zero dimensional model). With this simplification, we
argue that it reduces to the two matrix model and is indeed solvable.

We consider the partition function with the source term,
\begin{equation}
Z[J]=\int [d{a}] \exp\left( - \mbox{Tr}({a}\Lambda{a}) -\frac{1}{3} \mbox{Tr}%
({a}^3) -\mbox{Tr}(J{a}) \right)\,\,.
\end{equation}
The following change of variable,
\begin{equation}
{a}=-\Lambda+{a}^{\prime}
\end{equation}
kills the quadratic term and the partition function becomes,
\begin{equation}
Z[J]=e^{-\frac{2}{3}\mathrm{Tr}(\Lambda^3) + \mathrm{Tr} (J\Lambda)} \int [d{%
a}^{\prime}] \exp\left( -\frac{1}{3}\mbox{Tr}({{a}^{\prime}}^3) - \mbox{Tr}%
((J-\Lambda^2){a}^{\prime}) \right)\,\,.
\end{equation}
This is the partition function of the purely cubic theory with the modified
source term $J\rightarrow J^{\prime}=J-(\Lambda)^2$.

If we ignore the prefactor, the problem is now reduced to solve the
partition function,
\begin{equation}  \label{matrix2}
Z[J^{\prime}]\propto\int [d{a}^{\prime}] \exp\left( -\frac{1}{3}\mbox{Tr}({{a%
}^{\prime}}^3) - \mbox{Tr}( J^{\prime}{a}^{\prime}) \right)\,\,.
\end{equation}
One interesting aspect of this integration is that the off-diagonal part of
the matrix integration can be exactly performed. The measure of the
integration of Hermite matrix ${a}$ can be replaced by
\begin{equation}
[d{a}^{\prime}]=d\vec{a}\, [dU](\Delta(a))^2
\end{equation}
where we use the decomposition ${a}^{\prime}=Ua U^\dagger$ by using unitary
matrix $U$ and eigenvalues $\vec{a}$ of $A^{\prime}$ ($a=\mbox{diag}(\vec{a}%
) $). $\Delta(a)=\prod_{i<j}(a_i-a_j)$ is van der Monde determinant. After
this decomposition, Eq.(\ref{matrix2}) becomes,
\begin{equation}
\int d\vec{a}\, (\Delta(a))^2 e^{-\frac{1}{3}\sum_i a_i^3}\int[dU] \exp(-%
\mbox{Tr}J^{\prime}UaU^{-1})
\end{equation}
The integration over unitary matrix can be performed by using the famous
formula proved by Itzykson, Zuber and Brezin \cite{Mehta},
\begin{equation}
\int [dU] \exp\left(\frac{1}{t} \mbox{Tr}(AUBU^{-1})\right) = c
(\Delta(a)\Delta(b))^{-1} \det\left[ \exp(\frac{1}{t}a_jb_k) \right]\,\,,
\end{equation}
where $a,b$ are the eigenvalues of matrices $A,B$, $c=t^{N(N-1)/2}%
\prod_{j=1}^n j!$. Eq.(\ref{matrix2}) becomes finally,
\begin{equation}
Z[J^{\prime}]\propto \int d\vec{a}\, \frac{\Delta(a)}{\Delta{(\varphi)}} \,
\exp\left( -\frac{1}{3}\sum_{i} a_i^3 -\sum_{i} \varphi_i a_i \right)
\end{equation}
where $\varphi$'s are the eigenvalues of $J^{\prime}$.


\end{document}